\documentclass[3p,11pt]{elsarticle}
\usepackage{graphicx}
\usepackage{lineno}
\usepackage{comment}
\usepackage{hyperref}
\usepackage{color}
\usepackage[figuresleft]{rotating}
\usepackage{caption}
\usepackage{subcaption}
\usepackage{layout}
\usepackage{setspace}
\usepackage{dcolumn}
\usepackage{xtab}
\usepackage{tabularx}
\usepackage[T1]{fontenc} %together
\usepackage{tgtermes, tgheros} %together
\usepackage[scaled=0.92]{helvet}
\usepackage{amssymb} %disable this when mtpro2
\usepackage{autonum}
\usepackage{mathtools}
\usepackage{multirow}
\usepackage{amsfonts}
\usepackage{bm}

\newcommand{\Var}{\mathrm{Var}}

\usepackage{array}
\usepackage{tabularx}
\newcolumntype{C}{>{\centering\arraybackslash}X}
\newcolumntype{L}{>{\raggedright\arraybackslash}X}
\newcolumntype{R}{>{\raggedleft\arraybackslash}X}
\usepackage[table]{xcolor}
\newcommand*{\diff}{\mathrm{d}}
\journal{AAA}
\biboptions{authoryear}
\usepackage{hyperref}
\makeatletter
\def\ps@pprintTitle{%
\let\@oddhead\@empty
\let\@evenhead\@empty
\def\@oddfoot{}%
\let\@evenfoot\@oddfoot}
\makeatother
\usepackage{amsthm}
\newtheorem{propo}{Proposition}
\begin{document}
\onehalfspacing
\begin{frontmatter}
\title{ Productivity propagation with networks transformation}

%\author[sn]{First Author}
%\ead{fauthor@xxx.edu}
%\author[kn]{Second Author}
%\ead{sauthor@yyy.edu}
%\address[sn]{First Author's Affiliation and Address}
%\address[kn]{Second Author's Affiliation and Address}

\author[sn]{Satoshi Nakano}
\ead{snakano@n-fukushi.ac.jp}
\author[kn]{Kazuhiko Nishimura%\corref{cor1}
}
\ead{nishimura@lets.chukyo.ac.jp}
\address[sn]{Faculty of Economics, Nihon Fukushi University, Tokai Aichi 477-0031, Japan}
\address[kn]{School of Global Studies, Chukyo University, Nagoya Aichi 466-8666}
\date{\today}
\begin{abstract}
We model sectoral production by cascading binary compounding processes.
The sequence of processes is discovered in a self-similar hierarchical structure stylized in the economy-wide networks of production.
Nested substitution elasticities and Hicks-neutral productivity growth are measured such that the general equilibrium feedbacks between all sectoral unit cost functions replicate the transformation of networks observed as a set of two temporally distant input-output coefficient matrices.
We examine this system of unit cost functions to determine how idiosyncratic sectoral productivity shocks propagate into aggregate macroeconomic fluctuations in light of potential network transformation.
Additionally, we study how sectoral productivity increments propagate into the dynamic general equilibrium, thereby allowing network transformation and ultimately producing social benefits.
\end{abstract}
\begin{keyword}
Cascaded production \sep Total factor productivity \sep Restoring parameters \sep Dynamic general equilibrium \sep Nonlinearity and synergism
\end{keyword}
\end{frontmatter}
JEL Classification: E37, O33, O41
\section{Introduction}

Cobb-Douglas economy and frictionless factor substitution have been the key assumptions in multi-sector general equilibrium models examining %used to examine 
aggregate macroeconomic fluctuations.
The convenience of Cobb-Douglas assumption is that the (unit) substitution elasticity precisely offsets the price changes with a change in physical quantity, whereby the monetary input--output linkages %(i.e., networks) 
become fixed and linear throughout the propagation of the productivity shocks.
The works of \citet{lpJPE, horvath98, dupor, horvath} describe how sectoral productivity shocks propagate into aggregate macroeconomic fluctuations concerning the input-output linkages among multiple sectors.
More recenlty, while \citet{gabaixECTA} finds that the sum of independent random shocks with power law weights exhibit greater aggregate volatility than in the case with equal weights, 
\citet{aceECTA} discover that empirical input-output linkages lead to similar augmentation of aggregate volatility. 

Nevertheless, empirical analyses based on linked input-output tables do not support Cobb-Douglas, in majority of the disaggregated sectors \citep{knn}.\footnote{Linked input-output tables are time-series observations of intersectoral monetary transactions recorded in both nominal and real terms.} 
In this regard, one of the objectives of this study is to establish an alternative aggregator function with multiple factors, without imposing \textit{a priori} substitution elasticity.
Our empirical system of sectoral aggregator functions endogenizes factor substitutions, with sectoral Hicks-neutral productivity changes being the only exogenous force.
Whereas Cobb-Douglas aggregator is restrictive with respect to factor substitutions, our aggregator function is flexible to the extent that different production networks in two equilibrium states are replicable.  
By using our system of empirical aggregator functions, we can investigate the economy-wide propagation of sectoral productivity shocks, with explicit attention to the transformation of networks (input-output linkages) and the synergism that can take place by the cause of nonlinearity.

The aggregator function, which we refer to hereafter as the cascaded CES (CCES), consists of multiple binary CES (constant elasticity of substitution) processes, serially nested in a cascading manner.
Concerning the first-order conditions, the elasticity of substitution between the two factor inputs of a CRS (constant returns to scale) CES can be measured by regressing the log ratio of factor shares against the log ratio of factor prices \citep{acms, berndt, antras}.
The share parameter can be jointly specified as we normalize the data at the reference state.\footnote{Normalization of CES is extensively discussed in \citet{KdLG, normces}.}
We can then use the estimated parameters with the observed data to predict the compound output of the binary CES.
A two-stage CES production process utilizes this compound output as one of the two factor inputs, in addition to a third input.\footnote{
Log-linear regression of two-stage CRS CES aggregator functions is often applied in the empirical estimation of Armington elasticities.
Note, however, that previous studies do not use predicted values from the first-stage CES for the second-stage regression; instead, \citet{saito} uses the Laspeyres index; \citet{search} use the (multi-factor) Sato-Vartia index. }
Thence, the parameters of a CCES production function can be recursively estimated by a series of regression equations that utilize the predicted values of the compound output from the lower stage binary CES processes.

Two-stage CES production functions are often applied in econometric computable general equilibrium (CGE) models \citep[e.g.,][]{eneecon}, where the parameters are typically estimated by way of direct nonlinear regression, even under CRS assumption, leaving the first-order conditions unused.
Apart from CES, \citet{hudson} introduced translog production functions with four aggregated factors, now recognized as the precursor to the KLEM-type CGE models.
A second-order generalization of Cobb-Douglas, translog production functions are flexible with respect to substitution elasticities across the factor inputs.
When estimating the translog parameters, first-order conditions are typically utilized in addition to the main log-linear regression \citep{handbook}.
Either approach, however, requires a considerable number of time series observations of aggregated factor indices, and therefore, is generally difficult to extend the model to more than four disaggregated factors.

In contrast, CCES is capable of modeling production with many factor inputs.
The nested order of factor inputs enables the recursive system of regression equations to estimate the parameters of all binary CES processes.
In this regard, we use the sectoral hierarchy latent in the networks of production.
When each compound is viewed as an intermediate output, it compiles as the process compounds further, after which the input-output transactions become triangular at all levels of production.
This scale-freeness in a triangular input-output transaction is utilized to specify the intrasectoral sequence of the compounding processes (see \hyperlink{app1}{Appendix 1} for cascading order resolution).
In this study, CCES parameters are estimated via two-point regression, where the predictor completely restores the two observations of the regressand.
Thus, two equilibrium production networks are restored in the course of general equilibrium feedback within the system of empirical sectoral CCES unit cost functions with concomitant sectoral productivity changes.

We draw data from linked input-output tables \citep{miac} that have the following two-way (column-wise monetary and row-wise mass) balances in each time period $t=0, \cdots, T$.\footnote{As regards all measurements based on two-period observations in the subsequent sections (e.g., two-point regressions, index and growth calculations), we use the mean values of 2000 and 2005 as $t=0$, and those of 2005 and 2010 as $t=1$, from the three-point linked input-output tables \citet{miac}.}
\begin{align}
E_{jt}+ \sum_{i=1}^{I} p_{it} x_{ijt} = p_{jt} y_{jt}
&&
f_{it} + \sum_{j=1}^J x_{ijt} = y_{it}
\label{data}
\end{align}
There are $J$ sectors and $J=I$ intermediate goods, where the $i$th good is produced exclusively by the corresponding $i=j$th sector.
Value added, $E_j=rK_j + wL_j$, or the first compound, can be partitioned into two primary goods, i.e., capital service $rK_j$ and labor service $wL_j$.
Final demand, $f_i=h_i + g_i + m_i$, can be partitioned into household consumption, $h_i$, fixed capital formation, $g_i$, and net exports, $m_i$.
The cost share is defined as $s_{ij} = \frac{p_{i} x_{ij}}{p_{j} y_{j}}$ for all $i=1, \cdots, I$ and $j=1,\cdots,J$.
For primary goods, we note that ${s}_{0} = \frac{r K_{j}}{p_{j} y_{j}} = {s}_K$, and thus, $\frac{w L_{j}}{p_{j} y_{j}}=1-\sum_{i=0}^I s_{ij}={s}_L$.

%We hereafter refer to a square matrix $\mathbf{S}=(s_{ij})$ as a \textit{production network} and $\bm{S}=(\mathbf{S}, \bm{s}_K, \bm{s}_L)^\intercal$ as the cost share structure.

The remainder of the paper is organized as follows. 
The next section specifies the functional form of a CCES aggregator function and describes how the parameters can be estimated with respect to the first-order conditions. 
Section 3 introduces the model to replicate the empirical economy by CCES system, where all the parameters are obtained through two-point regression.
We show that the transition of production networks through the two points is endogenized by empirical sectoral productivity growth.
We then apply this model to study aggregate macroeconomic fluctuations in light of network transformation.
Section 4 provides a module for the representative household, estimating the indirect utility function in the form of multifactor CES, to evaluate productivity changes in terms of social benefits.
The nonlinearity of the non-Cobb-Douglas economy and the underlying synergism of productivity changes are discussed.
Section 5 provides concluding remarks.

\clearpage
\section{Production Economy}
\subsection{Two-stage CES production}
Below is a representation of a two-stage production function:
\begin{align}
\xi_2 
= F_2\left( x_1, \xi_1 \right) &&
\xi_1 = F_1\left( x_0, \xi_0 \right)   
\label{pf}
\end{align}
where $x_i$ and $\xi_n$ denote the physical quantities of the $i$th good and the $n$th compound that comes from the $n$th stage (process) of production, respectively.
Let us call the $i=0$th good and the $n=0$th compound primary factors, as there is no sector or process responsible for their supply.
The first stage produces the first compound $\xi_1$. 
In this production process there are $N=2$ stages and $I=N-1$ goods that are not primary (i.e., $x_1$), which we call intermediate.
By assuming CRS and CES for each stage of production, we have the following two-stage CES (dual) aggregator function:
\begin{align}
\pi_2 
&= c\left( p_1, \pi_1; \alpha_1, \gamma_1 \right)
=\left( \alpha_1 (p_1)^{\gamma_1} + (1-\alpha_1) (\pi_1)^{\gamma_1}\right)^{\frac{1}{\gamma_1}} 
\label{ucf2}
\\
\pi_1 
&= c\left( p_0, \pi_0; \alpha_0, \gamma_0  \right)
=\left( \alpha_0 (p_0)^{\gamma_0} + (1-\alpha_0) (\pi_0)^{\gamma_0}\right)^{\frac{1}{\gamma_0}} 
\label{ucf1}
\end{align}
where $p_i$ and $\pi_n$ denote the prices of the $i$th factor and the $n$th compound, respectively.
It will always be the case that the $i$th factor enters the $i+1=n$th stage.
For each stage $n=i+1$, $\alpha_i$ and $1-\gamma_i$ denote the share parameter and the elasticity of substitution between the $i$th factor and the $i$th compound, respectively.
Note that primary factor prices $p_0 = r$ and $\pi_0 = w$ and $p_1$ must be observable.

Because of the CRS assumption for (\ref{pf}), i.e., that $F_1$ and $F_2$ are homogeneous of degree one, the following zero profit condition must hold:
\begin{align}
\pi_2 \xi_2 = p_1 x_1 + \pi_1 \xi_1
&&
\pi_1 \xi_1 = p_0 x_0 + \pi_0 \xi_0
\label{zero}
\end{align}
By applying Shephard's lemma to the aggregator functions (\ref{ucf2}, \ref{ucf1}), i.e., 
\begin{align}
\frac{p_1}{\pi_2} \frac{\partial \pi_2}{\partial p_1} 
=\frac{p_1 x_1}{\pi_2 \xi_2}
\equiv {s}_1
&&
\frac{\pi_{1}}{\pi_2} \frac{\partial \pi_2}{\partial \pi_{1}} 
=\frac{\pi_1 \xi_1}{\pi_2 \xi_2}
&=1-{s}_1
\\
\frac{p_0}{\pi_2} \frac{\partial \pi_2}
{\partial p_0} 
=\frac{p_0 x_0}{\pi_2 \xi_2}
\equiv {s}_0
&&
\frac{\pi_{0}}{\pi_2} \frac{\partial \pi_2}
{\partial \pi_0} 
=\frac{\pi_0 \xi_0}{\pi_2 \xi_2}
&=1-{s}_0-{s}_1 
\label{sheph}
\end{align}
we have the following FOCs (first-order conditions):
\begin{align}
{s}_1 = \alpha_1 \left( \frac{p_1}{\pi_2} \right)^{\gamma_1}
&&
1- {s}_1 = (1-\alpha_1) \left( \frac{\pi_1}{\pi_2} \right)^{\gamma_1}
\label{stage2}
\\
\frac{{s}_0}{1-{s}_1}=\alpha_0 \left( \frac{p_0}{\pi_1} \right)^{\gamma_0}
&&
\frac{1- {s}_0 - {s}_1}{1-{s}_1}=(1-\alpha_0) \left( \frac{\pi_0}{\pi_1} \right)^{\gamma_0}
\label{stage1}
\end{align}
where $s_i$ denotes the cost share of the $i$th factor.
Regarding (\ref{zero}), the cost share is evaluated as ${s}_i =\frac{p_i x_i}{\pi_2 \xi_2}$ for $i=0,1$.
The cost share of the compound entering the $n$th stage is, therefore, $1 - \sum_{i=n-1}^I {s}_i$, where $I+1=N=2$ indicates the total number of stages.

Below is the simple regression equation concerning (\ref{stage1}) to estimate $\gamma_{0}$ by the slope and $\alpha_0$ by the intercept where sample observations are indexed by $t$, and $\epsilon_{1t}$ is the disturbance term:
\begin{align}
\ln \frac{{s}_{0t}}{1-{s}_{0t} - {s}_{1t}} 
&= \ln \frac{\alpha_{0}}{1-\alpha_{0}} + \gamma_{0} \ln \frac{p_{0t}}{\pi_{0t}} + \epsilon_{1t}
\label{reg1}
\end{align}
On the other hand, to estimate $\gamma_1$ and $\alpha_1$, we need data for $\pi_1$, concerning (\ref{stage2}), although they are not observable.
To estimate these parameters, we use the predicted values $\hat{\pi}_{1t}$ obtained from the lower stage CES aggregator function using estimated parameters $\left( \hat{\gamma}_0, \hat{\alpha}_0 \right)$ from (\ref{reg1}).
That is,
\begin{align}
\hat{\pi}_{1t} 
=\left( \hat{\alpha}_0 (p_{0t})^{\hat{\gamma}_{0}} + (1-\hat{\alpha}_0) (\pi_{0t})^{\hat{\gamma}_{0}}\right)^{{1}/{\hat{\gamma}_{0}}} 
\label{hatpi1}
\end{align}
Then, the second-stage regression equation becomes the following:
\begin{align}
\ln \frac{{s}_{1t}}{1-{s}_{1t}} &= \ln \frac{\alpha_{1}}{1-\alpha_{1}} + \gamma_{1} \ln \frac{p_{1t}}{\hat{\pi}_{1t}} + \epsilon_{2t}
\label{reg2}
\end{align}
As we continue the procedure to predict $\hat{\pi}_{2t}$ using the estimated parameters $\left( \hat{\gamma}_1, \hat{\alpha}_1 \right)$ from (\ref{reg2}), we have:
\begin{align}
\hat{\pi}_{2t} 
=\left( \hat{\alpha}_1 (p_{1t})^{\hat{\gamma}_{1}} + (1-\hat{\alpha}_1) (\hat{\pi}_{1t})^{\hat{\gamma}_{1}}\right)^{{1}/{\hat{\gamma}_{1}}} 
\label{hatpi2}
\end{align}
If the empirical (two-stage) CES aggregator function is to be normalized at $t=1$, where $p_{01}=p_{11}=\pi_{01}=\pi_{11}=1$, the share parameters must be $\alpha_{1}={s}_{11}$ and $\alpha_{0}= \frac{{s}_{01}}{1-{s}_{11}}$, with respect to (\ref{reg1}) and (\ref{reg2}).
Note that whereas the estimated parameters $\left( \hat{\gamma}_i, \hat{\alpha}_i \right)$ are obtained through independent minimization of the SSRs of the first (\ref{reg1}) and the second (\ref{reg2}) regression, alternative estimates 
can be obtained through joint minimization of the two SSRs, i.e., $\min_{\gamma_0, \alpha_0, \gamma_1, \alpha_1} \sum_{t=0}^T \left( \epsilon_{1t} \right)^2 + \left( \epsilon_{2t} \right)^2$.
Since the joint minimization nests independent minimizations, the overall fit 
must be better under the joint minimization policy.

Let us now suppose that we have observations for two points $t=0,1$.
In this case the following proposition holds true.
\begin{propo}\label{propo1}
The two temporally distant cost shares of factors are restored as FOCs of the two-point estimated two-stage CES aggregator function.
\begin{proof}
In the case of two-point regression, estimators create null error terms, as regards (\ref{reg1}, \ref{reg2}), i.e.,
\begin{align}
\ln \frac{{s}_{0t}}{1-{s}_{0t} - {s}_{1t}} - \ln \frac{\hat{\alpha}_{0}}{1-\hat{\alpha}_{0}} - \hat{\gamma}_{0} \ln \frac{p_{0t}}{\pi_{0t}} 
= 0 && t=0,1 \label{one}
\\
\ln \frac{{s}_{1t}}{1-{s}_{1t}} - \ln \frac{\hat{\alpha}_{1}}{1-\hat{\alpha}_{1}} - \hat{\gamma}_{1} \ln \frac{p_{1t}}{\hat{\pi}_{1t}} 
= 0 && t=0,1 \label{two}
\end{align}
Under the two-point estimated parameters $(\hat{\alpha}_0, \hat{\alpha}_1, \hat{\gamma}_0, \hat{\gamma}_1)$, stage-wise aggregator functions are evaluated as follows:
\begin{align}
\hat{\pi}_{1t} 
=\left( \hat{\alpha}_0 (p_{0t})^{\hat{\gamma}_{0}} + (1-\hat{\alpha}_0) (\pi_{0t})^{\hat{\gamma}_{0}}\right)^{{1}/{\hat{\gamma}_{0}}} 
&& t=0,1  \label{three}
\\
\hat{\pi}_{2t} 
=\left( \hat{\alpha}_1 (p_{1t})^{\hat{\gamma}_{1}} + (1-\hat{\alpha}_1) (\hat{\pi}_{1t})^{\hat{\gamma}_{1}}\right)^{{1}/{\hat{\gamma}_{1}}} 
&& t=0,1 \label{four}
\end{align}
We may solve (\ref{one}, \ref{two}) for the two-point estimated parameters as follows:
\begin{align}
\hat{\gamma}_{0} &= \frac{ \ln \frac{{s}_{01}}{1-{s}_{01}-{s}_{11}}  - \ln \frac{{s}_{00}}{1-{s}_{00}-{s}_{10}} }
{\ln \frac{p_{01}}{{\pi}_{01}} - \ln \frac{p_{00}}{{\pi}_{00}}}
&
\ln \frac{\hat{\alpha}_0}{1-\hat{\alpha}_0} 
&=
\frac{\ln \frac{p_{01}}{{\pi}_{01}} \ln \frac{{s}_{00}}{1-{s}_{00} -{s}_{10}}  - \ln \frac{p_{00}}{{\pi}_{00}} \ln \frac{{s}_{01}}{1- {s}_{01} - {s}_{00}} }{\ln \frac{p_{01}}{{\pi}_{01}} - \ln \frac{p_{00}}{{\pi}_{00}}}
\\
\hat{\gamma}_{1} &=
\frac{ \ln \frac{{s}_{11}}{1-{s}_{11}}  - \ln \frac{{s}_{10}}{1-{s}_{10}} }
{\ln \frac{p_{11}}{\hat{\pi}_{11}} - \ln \frac{p_{10}}{\hat{\pi}_{10}}}
&
\ln \frac{\hat{\alpha}_1}{1-\hat{\alpha}_1} 
&=
\frac{\ln \frac{p_{11}}{{\pi}_{11}} \ln \frac{{s}_{10}}{1-{s}_{10}}  - \ln \frac{p_{10}}{{\pi}_{10}} \ln \frac{{s}_{11}}{1- {s}_{11}} }{\ln \frac{p_{11}}{\hat{\pi}_{11}} - \ln \frac{p_{10}}{\hat{\pi}_{10}}}
\end{align}
Thus, with the help of (\ref{three}), not only $\hat{\gamma}_0$ and $\hat{\alpha}_0$, but $\hat{\gamma}_1$ and $\hat{\alpha}_1$ can also be solved. 
Furthermore, we may solve for the observed cost shares $(s_{10}, s_{11}, s_{00}, s_{01})$, from (\ref{one}, \ref{two}, \ref{three}, \ref{four}), as follows: 
\begin{align}
s_{1t}=\hat{\alpha}_1 \left( \frac{p_{1t}}{\hat{\pi}_{2t}} \right)^{\hat{\gamma}_1}
&&
1- s_{1t}=(1-\hat{\alpha}_1) \left( \frac{\hat{\pi}_{1t}}{\hat{\pi}_{2t}} \right)^{\hat{\gamma}_1}
&&
t=0,1
\label{foca}
\\
\frac{s_{0t}}{1-s_{1t}}= \hat{\alpha}_0 \left( \frac{p_{0t}}{\hat{\pi}_{1t}} \right)^{\hat{\gamma}_0}
&&
\frac{1-s_{0t}-s_{1t}}{1-s_{1t}}= (1-\hat{\alpha}_0 )\left( \frac{\pi_{0t}}{\hat{\pi}_{1t}} \right)^{\hat{\gamma}_0}
&&
t=0,1
\label{focb}
\end{align}
In regard to (\ref{stage2}, \ref{stage1}), we know that (\ref{foca}, \ref{focb}) are the FOCs for the two-point estimated two-stage CES aggregator function.
\end{proof}
\end{propo}

For later convenience let us work on the two-point estimated two-stage CES aggregator function $C$, i.e.,
\begin{align}
\pi_2 &=
C( p_1, p_0, \pi_0 )
\equiv
c\left( p_1, \pi_1:= c\left( p_0, \pi_0; \hat{\alpha}_0, \hat{\gamma}_0 \right); \hat{\alpha}_1, \hat{\gamma}_1\right)
\\
&=\left( \hat{\alpha}_1 (p_{1})^{\hat{\gamma}_1} + (1 - \hat{\alpha}_1) \left(  
{\pi_1:=
\left( \hat{\alpha}_0 (p_{0})^{\hat{\gamma}_0} + (1 - \hat{\alpha}_0) \left( \pi_0 \right)^{\hat{\gamma}_0}  \right)^{1/\hat{\gamma}_0}
}
 \right)^{\hat{\gamma}_1}  \right)^{1/\hat{\gamma}_1}
\end{align}
By partially differentiating $C$, we obtain the following FOCs:
\begin{align}
\frac{p_1}{\pi_2}\frac{\partial C}{\partial p_1} &= \hat{\alpha}_1 \left( \frac{p_1}{\pi_2} \right)^{\hat{\gamma}_1}
&
\frac{\pi_1}{\pi_2}\frac{\partial C}{\partial \pi_1} &=(1- \hat{\alpha}_1) \left( \frac{\pi_1}{\pi_2} \right)^{\hat{\gamma}_1}
\\
\frac{p_0}{\pi_2}\frac{\partial C}{\partial p_0} &= \hat{\alpha}_0 (1-\hat{\alpha}_1) \left( \frac{p_0}{\pi_1} \right)^{\hat{\gamma}_0} \left( \frac{\pi_1}{\pi_2} \right)^{\hat{\gamma}_1}
&
\frac{\pi_0}{\pi_2}\frac{\partial C}{\partial \pi_0} &= (1-\hat{\alpha}_0) (1-\hat{\alpha}_1) \left( \frac{\pi_0}{\pi_1} \right)^{\hat{\gamma}_0} \left( \frac{\pi_1}{\pi_2} \right)^{\hat{\gamma}_1}
\end{align}
Thus, evaluation of the FOCs for the two periods (points) $t=0,1$, yields (\ref{foca}, \ref{focb}) which may be described concisely using $C$ as follows:
\begin{align}
\frac{\left< p_{1t}, p_{0t}, \pi_{0t} \right> \nabla C(p_{1t}, p_{01}, \pi_{0t})}{C(p_{1t}, p_{01}, \pi_{0t})}
= 
\left( s_{1t}, s_{0t}, 1- s_{0t} - s_{1t}\right)^\intercal 
&&
t=0,1
\label{ppp}
\end{align}
where $\nabla$ indicate a partial derivative with respect to each argument (gradient operator), and angle brackets indicate diagonalization of a vector. 
Note that (\ref{ppp}) and Proposition \ref{propo1} are equivalent.

Let us now consider the index number that corresponds exactly to the aggregator function $C$.
By definition we know that (log-form) index $\Delta \ln \hat{\pi}_{2t} = \ln (\hat{\pi}_{21}/\hat{\pi}_{20})$ is exact \citep{diewert} for $C$, i.e.,
\begin{align}
\Delta \ln \hat{\pi}_{2t} = \Delta \ln {C(p_{1t}, p_{0t}, \pi_{0t})}
\end{align}
By the following proposition we know that index $\Delta \ln \hat{\pi}_{2t}$ can be evaluated by the change in prices and an optimal response in quantities (i.e., change in cost shares of factors).
\begin{propo} \label{coro1}
A two-stage Sato-Vartia index is exact for a two-point estimated two-stage CES aggregator function.
\begin{proof}
It must suffice to show that $\Delta \ln \hat{\pi}_{2t}$ is equal to the two-stage two-factor Sato-Vartia index.
By taking the log and temporally difference of (\ref{focb}) we have:
\begin{align}
\Delta \ln {\phi}_{0t} \equiv
\Delta \ln \frac{s_{0t}}{1-s_{1t}}
=\hat{\gamma}_0 \Delta \ln \frac{p_{0t}}{\hat{\pi}_{1t}}
&&
\Delta \ln {\phi}_{0t}^\prime \equiv
\Delta \ln \frac{1-s_{0t}-s_{1t}}{1-s_{1t}}
=\hat{\gamma}_0 \Delta \ln \frac{\pi_{0t}}{\hat{\pi}_{1t}}
\label{svone}
\end{align}
where ${\phi}_{0t}+{\phi}^\prime_{0t}=1$ for $t=0,1$ by definition.
Elimination of $\hat{\gamma}_{0}$ yields,
\begin{align}
\Delta \ln \hat{\pi}_{1t}=
\frac{\Delta \ln \pi_{0t}\Delta \ln {\phi}_{0t} - \Delta \ln p_{0t}\Delta\ln {\phi}_{0t}^\prime}{\Delta \ln {\phi}_{0t} - \Delta\ln {\phi}_{0t}^\prime}
\end{align}
This identity is reduced into the following form, using $\Delta {\phi}_{0t}+ \Delta {\phi}_{0t}^\prime =0$.
\begin{align}
\Delta \ln \hat{\pi}_{1t}
&=
\frac{\Delta \ln \pi_{0t}\Delta \ln {\phi}_{0t} \Delta {\phi}_{0t}^\prime + \Delta \ln p_{0t}\Delta\ln {\phi}_{0t}^\prime \Delta {\phi}_{0t}}{\Delta \ln {\phi}_{0t}\Delta {\phi}_{0t}^\prime + \Delta\ln {\phi}_{0t}^\prime \Delta {\phi}_{0t}}
\\
&=
\frac{\Delta \ln \pi_{0t}\Delta \ln \frac{s_{0t}}{1-s_{1t}} \Delta \frac{1-s_{0t}-s_{1t}}{1-s_{1t}} + \Delta \ln p_{0t}\Delta\ln \frac{1-s_{0t}-s_{1t}}{1-s_{1t}} \Delta \frac{s_{0t}}{1-s_{1t}}}{\Delta \ln \frac{s_{0t}}{1-s_{1t}}\Delta \frac{1-s_{0t}-s_{1t}}{1-s_{1t}} + \Delta\ln \frac{1-s_{0t}-s_{1t}}{1-s_{1t}} \Delta \frac{s_{0t}}{1-s_{1t}}}
\label{sv1}
\end{align}
Note that this is the two-factor Sato-Vartia index \citep{sato_restat, vartia} for the first stage aggregator.\footnote{While a two-factor Sato-Vartia index corresponds exactly to a two-factor CES aggregator function, a multi-factor Sato-Vartia index does not correspond exactly to a multi-factor CES aggregator function \citep{lau}.}
Further, by taking the log and temporally difference of (\ref{foca}) we obtain: 
\begin{align}
\Delta \ln {\phi}_{1t} \equiv
\Delta \ln{s_{1t}}
=\hat{\gamma}_1 \Delta \ln \frac{p_{1t}}{\hat{\pi}_{2t}}
&&
\Delta \ln {\phi}_{1t}^\prime \equiv
\Delta \ln ({1-s_{1t}})
=\hat{\gamma}_1 \Delta \ln \frac{\pi_{1t}}{\hat{\pi}_{2t}}
\label{svtwo}
\end{align}
where ${\phi}_{1t}+{\phi}^\prime_{1t}=1$ for $t=0,1$ by definition.
Elimination of $\hat{\gamma}_{1}$ yields,
\begin{align}
\Delta \ln \hat{\pi}_{2t}=
\frac{\Delta \ln \hat{\pi}_{1t}\Delta \ln {\phi}_{1t} - \Delta \ln p_{1t}\Delta\ln {\phi}_{1t}^\prime}{\Delta \ln {\phi}_{1t} - \Delta\ln {\phi}_{0t}^\prime}
\end{align}
This identity is reduced into the following form, using $\Delta {\phi}_{1t}+ \Delta {\phi}_{1t}^\prime =0$.
\begin{align}
\Delta \ln \hat{\pi}_{2t}
&=
\frac{\Delta \ln \hat{\pi}_{1t}\Delta \ln {\phi}_{1t} \Delta {\phi}_{1t}^\prime + \Delta \ln p_{1t}\Delta\ln {\phi}_{1t}^\prime \Delta {\phi}_{1t}}{\Delta \ln {\phi}_{1t}\Delta {\phi}_{1t}^\prime + \Delta\ln {\phi}_{1t}^\prime \Delta {\phi}_{1t}}
\\
&=
\frac{\Delta \ln \hat{\pi}_{1t}\Delta \ln s_{1t} \Delta (1-s_{1t}) + \Delta \ln p_{1t}\Delta \ln (1-s_{1t}) \Delta s_{1t} }{\Delta \ln s_{1t} \Delta (1-s_{1t}) + \Delta\ln (1-s_{1t}) \Delta s_{1t}}
\label{sv2}
\end{align}
Combining (\ref{sv1}, \ref{sv2}) yields the two-stage Sato-Vartia index. 
\end{proof}
\end{propo}

\subsection{Cascaded ($N$-stage) CES production}
CCES production is a simple extension of two-stage CES production.
Regarding the dimension of our empirical model of the production economy, there are $I=385$ intermediate goods that are produced by $J=385$ corresponding sectors.
There are thus (at most) $N=I+1=386$ binary process stages in a sector's production.
The primary stage ($n=1$) for all sectoral production processes aggregates two primary factors $x_0=K$ and $\xi_0=L$, with corresponding prices, $p_0=r$ and $\pi_0=w$.  
We estimate the parameters of our model based on a set of empirical linked input--output tables whose $J$ sectors are ordered following Colin Clark's three-sector theory, which we call the \textit{classification order}.
The subsequent estimation procedure follows the \textit{cascading order} of sectors that reflects the downstreaming nature of intrasectoral binary processes uncovered by triangulating the empirical input-output incidence matrix. 
See \hyperlink{app1}{Appendix 1} for details.

Corresponding to (\ref{stage2}, \ref{stage1}) 
the first-order conditions for a multi-stage CES aggregator function are:
\begin{align}
\frac{s_{i}}{1-\sum_{k=i+1}^I {s_{k}}} = \alpha_{i} \left( \frac{p_{i}}{\pi_{n}} \right)^{\gamma_{i}}
&&
\frac{1-\sum_{k=i}^I s_{k}}{1-\sum_{k=i+1}^I {s_{k}}} = (1-\alpha_{i}) \left( \frac{\pi_{i}}{\pi_{n}} \right)^{\gamma_{i}}
\label{focm}
\end{align}
and the regression equation for the $i+1 =n$th nest (of the cascading order), processing the $i$th factor and the $i$th compound can be written as follows:
\begin{align}
\ln \frac{{s}_{it}}{1-\sum_{k=i}^I {s}_{k t}} 
= \ln \frac{\alpha_i}{1-\alpha_i} + \gamma_{i} \ln \frac{p_{it}}{\hat{\pi}_{it}} +\epsilon_{nt}
&&
n=1,\cdots,N
\label{object}
\end{align}
where $\hat{\pi}_{0t}= \pi_{0t}$ for $n=1$.
The $n$th compound price $\hat{\pi}_{n} =\hat{\pi}_{i+1}$ can be evaluated by the following predictors: 
\begin{align}
\hat{\pi}_{n t} 
= \left( \hat{\alpha}_{i} (p_{it})^{\hat{\gamma}_i} + (1-\hat{\alpha}_{i}) (\hat{\pi}_{it})^{\hat{\gamma}_i} \right)^{{1}/{\hat{\gamma}_i}}
&&
n=1,\cdots,N
\label{update}
\end{align}
When estimating the parameters, the sum of the nested SSR is minimized, i.e., 
\begin{align}
\left( \hat{\gamma}_{i}, \hat{\alpha}_{i} \right)=\arg \min_{\gamma_{i}, \alpha_{i}} \sum_{n=1}^{N} \sum_{t=0}^T \left( \epsilon_{nt} \right)^{2}
\text{  subject to  (\ref{object}) and (\ref{update}) }
\label{nlp}
\end{align}
This is a nonlinear programming (NLP) problem that can be solved by a nonlinear optimizer.
Alternatively, the problem can be viewed as a dynamic control problem with $(\gamma_{i}, \alpha_{i})$ being the control, and $\hat{\pi}_{nt}$ being the state (of $T$ dimension), which is updated $n$-wise by (\ref{update}).

As regards the two-point regression under minimum temporal observations $t=0,1$, the following propositions hold true.
\begin{propo}\label{propon}
The two temporally distant cost shares of factors are restored as FOCs of the two-point estimated $N$-stage CCES aggregator function.
\end{propo}
\begin{propo} \label{coro2}
An $N$-stage Sato-Vartia index is exact for a two-point estimated $N$-stage CCES aggregator function.
\end{propo}
Since these propositions are simple generalizations of Propositions \ref{propo1} and \ref{coro1}, respectively, the proofs are straightforward and therefore omitted.
Below, we display the two-point estimated parameters for a CCES aggregator function:
\begin{align}
\hat{\gamma}_{i} &= \frac{ \ln \frac{{s}_{i1}}{1-\sum_{k=i}^I {s}_{k1}}  - \ln \frac{{s}_{i0}}{1-\sum_{k=i}^I {s}_{k0}} }{\ln \frac{p_{i1}}{\hat{\pi}_{i1}} - \ln \frac{p_{i0}}{\hat{\pi}_{i0}}}
&i=0,1,\cdots,I \label{gammai}
\\
\ln \frac{\hat{\alpha}_i}{1-\hat{\alpha}_i} &= \frac{\ln \frac{p_{i1}}{\hat{\pi}_{i1}} \ln \frac{{s}_{i0}}{1-\sum_{k=i}^I {s}_{k0}}  - \ln \frac{p_{i0}}{\hat{\pi}_{i0}} \ln \frac{{s}_{i1}}{1-\sum_{k=i}^I {s}_{k1}} }{\ln \frac{p_{i1}}{\hat{\pi}_{i1}} - \ln \frac{p_{i0}}{\hat{\pi}_{i0}}}
&i=0,1,\cdots,I
\label{alphai}
\end{align}
where $\hat{\pi}_{0t}=\pi_{0t}$ for $t=0,1$.
The parameters $(\hat{\alpha}_{i}, \hat{\gamma}_{i})$ for $i=0,1,\cdots,I$ are to be recursively solved using $(\hat{\pi}_{i0}, \hat{\pi}_{i1})$, which will be evaluated by the two-point estimated $i+1=n$th stage CES aggregator function (\ref{update}). 
As regards Proposition \ref{coro2}, the $i+1=n$th stage Sato-Vartia index for an $I+1=N$ stage CCES aggregator function can be recursively obtained by the following formula:
\begin{align}
\Delta \ln \hat{\pi}_{nt}
=
\frac{\Delta \ln \hat{\pi}_{it}\Delta \ln {\phi}_{it} \Delta {\phi}_{it}^\prime + \Delta \ln p_{it}\Delta\ln {\phi}_{it}^\prime \Delta {\phi}_{it}}{\Delta \ln {\phi}_{it}\Delta {\phi}_{it}^\prime + \Delta\ln {\phi}_{it}^\prime \Delta {\phi}_{1t}}
\label{csvn}
\end{align}
where, 
%$\phi_{it} + \phi_{it}^\prime =1$ for $t=0,1$, %and thus $\Delta \phi_{it} + \Delta \phi_{it}^\prime =0$, 
%as we set, 
in light of (\ref{focm}), we use,
\begin{align}
\phi_{it} = \frac{s_{it}}{1-\sum_{k=i+1}^I s_{kt}}
&&
\phi_{it}^\prime = \frac{1-\sum_{k=i}^I s_{kt}}{1-\sum_{k=i+1}^I s_{kt}}
\label{csvnphi}
\end{align}

Concerning Proposition \ref{propon}, we describe below the FOCs of the two-point estimated $N$-stage CCES aggregator function $C(p_I, p_{I-1}, \cdots, p_1, p_0, \pi_0; \hat{\alpha}_0, \hat{\alpha}_1, \cdots, \hat{\alpha}_I, \hat{\gamma}_0, \hat{\gamma}_1, \cdots, \hat{\gamma}_I) \equiv C(\bm{p}, r, w)$, as follows:
\begin{align}
\frac{\left< \bm{p}_{t}, r_{t}, w_{t} \right> \nabla C(\bm{p}_{t}, r_{t}, w_{t})}{C(\bm{p}_{t}, r_{t}, w_{t})} = \left( \bm{s}_{t}, s_{Kt}, s_{Lt} \right)^\intercal && t=0,1
\label{pppn}
\end{align}
where $\bm{s}_t = (s_I, s_{I-1}, \cdots, s_1)$, $s_{K}=s_0$ and $s_{L}=1-\sum_{i=0}^I s_{i}$.
Note that (\ref{pppn}) is a simple extension of (\ref{ppp}) of the two-stage (three-factor) case.
Further let us introduce $\hat{\tau}_t$ by the following definition:
\begin{align}
q_t \hat{\tau}_t = \hat{\pi}_{Nt} =C(\bm{p}_t, r_t, w_t) && t=0,1
\label{qtaun}
\end{align}
where $q_t$ denotes the observed output price (unit cost) in period $t$.
Then we know by Proposition \ref{coro2} that $\Delta \ln \hat{\tau}_t$ is the total factor productivity growth (TFPg) for the two-point estimated $N$-stage CCES aggregator function that is evaluable by the $N$-stage Sato-Vartia (log-form) index $\Delta \ln \hat{\pi}_{Nt}$, via (\ref{csvn}, \ref{csvnphi}), i.e.,
\begin{align}
\text{TFPg (CCES)}=
\Delta \ln \hat{\tau}_t = \Delta \ln C(\bm{p}_t, r_t, w_t) - \Delta \ln q_t
= \Delta \ln \hat{\pi}_{Nt} - \Delta \ln q_t 
\end{align}

\begin{figure}[t!]
\centering
    \includegraphics[width=0.495\textwidth]{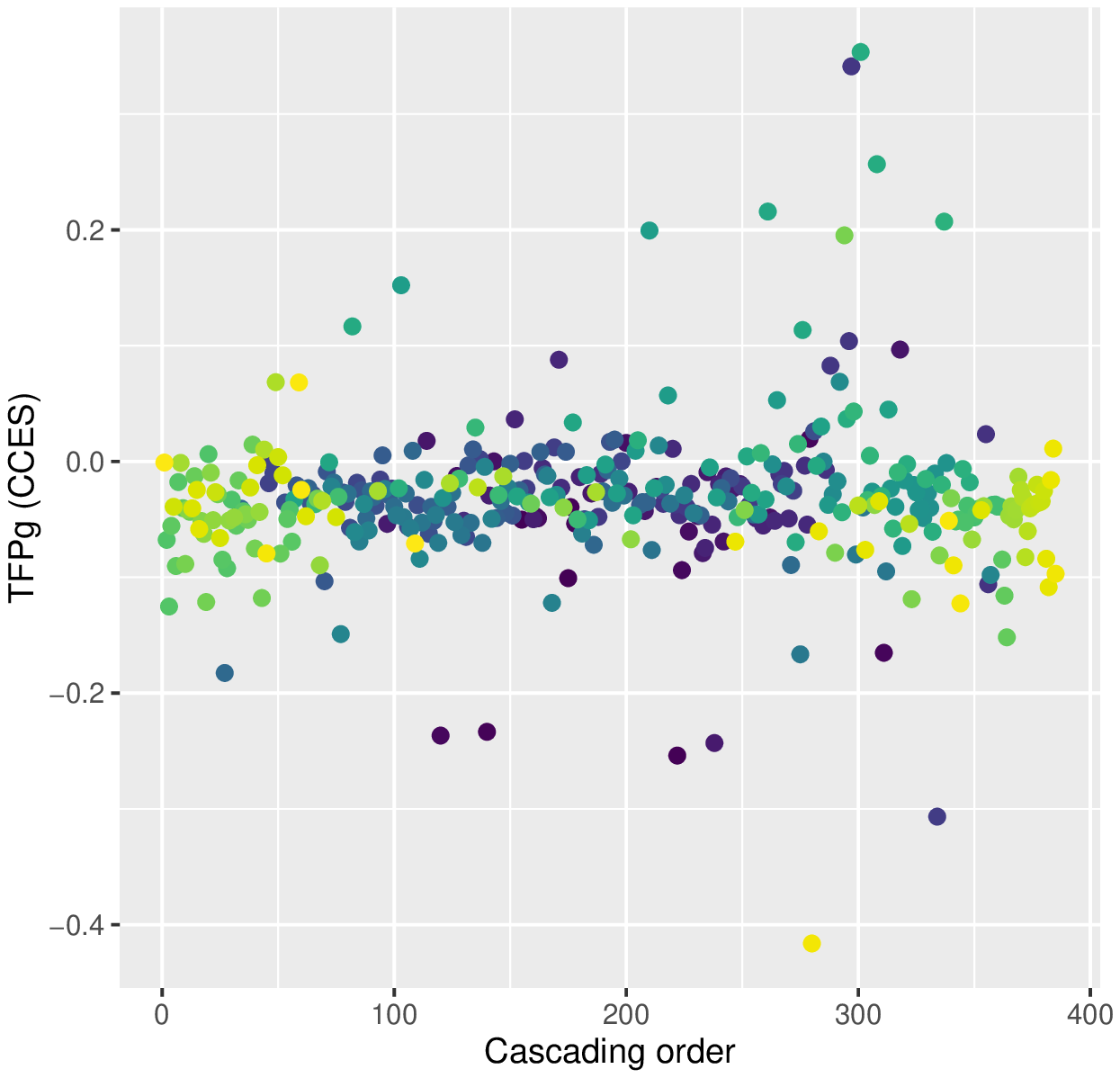}
    \includegraphics[width=0.495\textwidth]{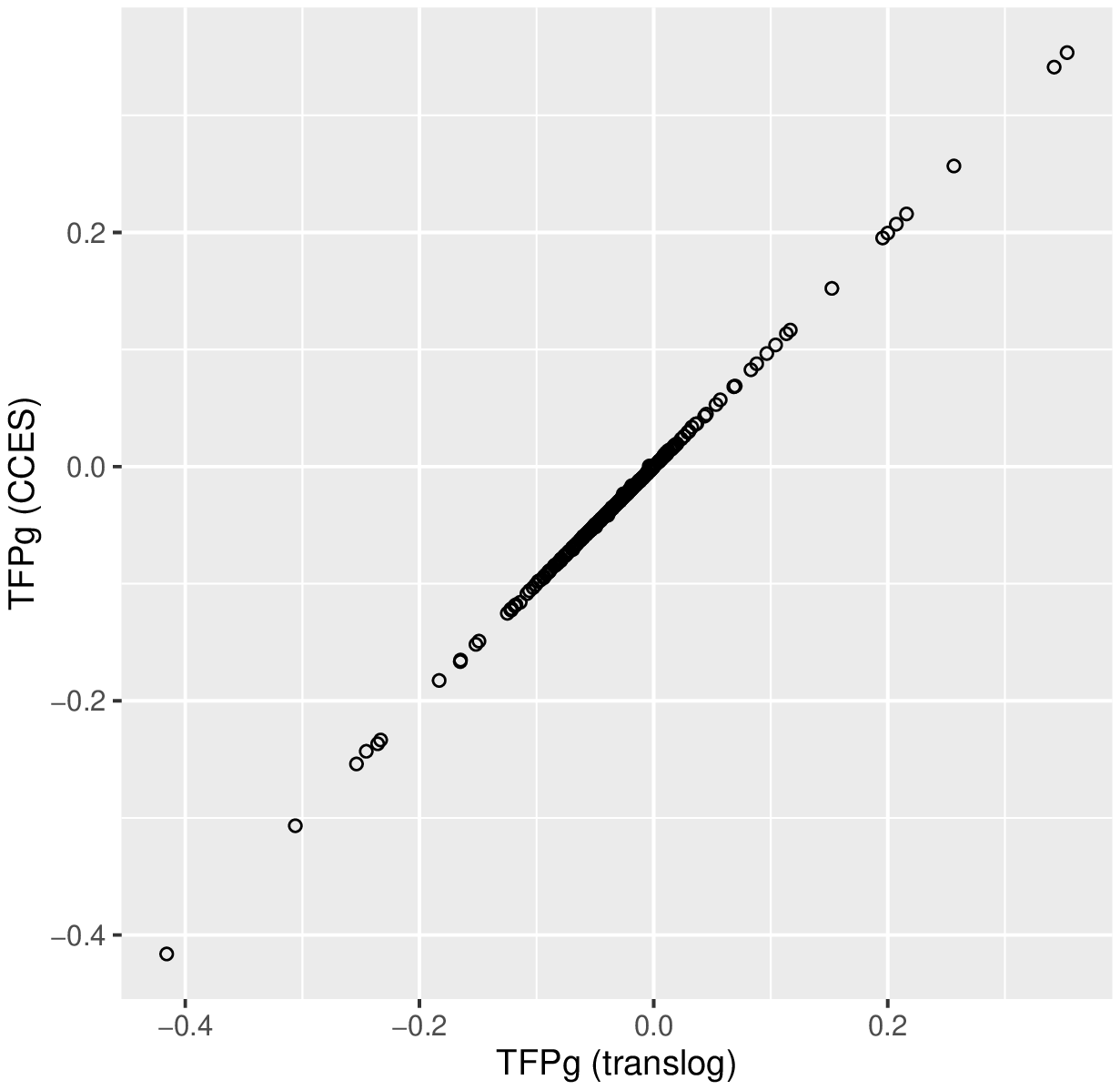}
    \caption{Left: Sectoral TFPg measurement based on CCES with two-point estimated (restoring) parameters. Colors correspond to the classification order of sectors. 
Right: Correspondences between TFPg (CCES) and TFPg (translog). All calculations are based on the linked input-output tables \citet{miac}.}
    \label{fig_tfpg}
\end{figure}

Concerning general equilibrium, the output price for all sectors must coincide with the corresponding intermediate good's price, i.e., $q_{jt} = p_{it}$ for $i=j=1, \cdots, I=J$ and $t=0,1$.
Thereupon, Figure \ref{fig_tfpg} (left) displays TFPg (CCES) for all $J=I$ sectors.
As for reference, we consider the following TFPg based on T{\"o}rnqvist (log-form) index, labeled as TFPg (translog).
Note that T{\"o}rnqvist index is exact for the underlying (cost-share-restoring) translog aggregator function \citep{diewert}.
\begin{align}
\text{TFPg (translog)} = \sum_{i=1}^I \bar{s}_{i} \Delta \ln {p_{it}}
+ \bar{s}_{K} \Delta \ln r_{t} 
+ \bar{s}_{L} \Delta \ln w_{t}
- \Delta \ln {q_{t}}
\end{align}
Here, we denote $\bar{s}_i=\left( {s}_{i1} + {s}_{i0} \right)/2$ for $i=1, \cdots, I, K, L$.  
Figure \ref{fig_tfpg} (right) displays TFPg (CCES) and TFPg (translog) for all sectors $j=1,\cdots,J$, showing extreme concordances between the two measurements.

\section{Production Networks}
\subsection{Networks Transformation}
Let us hereafter call the CCES aggregator function with certain estimated parameters as \textit{empirical}, and the two-point estimated parameters as \textit{restoring}.
As regards (\ref{qtaun}), an empirical CCES unit cost function for a sector (index omitted) can be described as follows:
\begin{align}
q
= \tau^{-1} C(\bm{p}, r, w)
= \tau^{-1} c(p_I, \cdots, c(p_1, c(r,w; \hat{\alpha}_0, \hat{\gamma}_0); \hat{\alpha}_1, \hat{\gamma}_1);\cdots;\hat{\alpha}_I, \hat{\gamma}_I )
\label{ccesucf}
\end{align} 
where $q$ denotes the unit cost, dependent upon the price of intermediate inputs $\bm{p}=(p_1, \cdots, p_I)$, and primary inputs $(r, w)$, and the level of productivity $\tau$.
The economy-wide system of empirical sectoral CCES unit cost functions is thus,
\begin{align}
\bm{q}=\bm{C}( \bm{p}, r, w ) \left< {\bm{\tau}} \right>^{-1}= \left( ({\tau}_{1})^{-1}C_1( \bm{p}, r, w), \cdots, ({\tau}_{J})^{-1}C_J( \bm{p}, r, w) \right)
\end{align}
where $\bm{q}=\left( q_1,\cdots, q_J \right)$ and $\bm{\tau} = \left( \tau_1,\cdots, \tau_J \right)$.
Note that the subscript $j=1,\cdots, J$ indicates the sector producing the $j$th commodity, exclusively.

Let us then consider a mapping $\mathcal{E}: \left(\bm{\tau}; r,w \right) \to \bm{p}$ according to the following system of empirical CCES unit cost functions providing all (intermediate) commodity prices:
\begin{align}
\bm{p}=\bm{C} \left( \bm{p}, r, w  \right) \left< {\bm{\tau}}\right>^{-1}
\label{emm}
\end{align}
The mapping $\mathcal{E}$ nests fix-point calculation of a system of nonlinear (presumably concave) functions $\bm{C}$, the fixed point of which is solvable by contractive feedback of $\bm{p}$ \citep{kras, kennan}.
That is, for a given $\bm{\tau}$, $\bm{p}$ is determined through {general equilibrium feedback iteration}.
Moreover, if the empirical CCES aggregator function is restoring, the two equilibrium states must be installed in (\ref{emm}), i.e.,  
\begin{align}
\bm{p}_t=\bm{C}( \bm{p}_t, r_t, w_t ) \left< \hat{\bm{\tau}}_t \right>^{-1}
&&
t=0,1
\label{genn}
\end{align}
This is the multi-sector version of (\ref{qtaun}) with two equilibrium states $t=0,1$ where $\bm{q}_t = \bm{p}_t$.
Applying Shephard's lemma, the equilibrium cost-share structure can be derived by the gradient of (\ref{emm}), i.e.,
\begin{align}
\left< \bm{p}, r, w \right> \nabla \bm{C}\left( \bm{p}, r, w  \right)
\left< \bm{\tau} \right>^{-1} \left< \bm{p} \right>^{-1}
=\bm{S}
\label{focnn}
\end{align}
where $\bm{S}=\left( \mathbf{S}, \bm{s}_K, \bm{s}_L \right)^\intercal$ denotes an $(I+2)\times J$ matrix, element ${s}_{ij}$ of which denotes the cost share of the $i$th factor of the $j$th sector. 
We refer to $\bm{S}$ as production networks.
If the aggregator function is restoring, the two equilibrium states must be installed in (\ref{focnn}), i.e., 
\begin{align}
\left< \bm{p}_t, r_t, w_t \right> \nabla \bm{C}\left( \bm{p}_t, r_t, w_t  \right)
\left< \hat{\bm{\tau}}_t \right>^{-1} \left< \bm{p}_t \right>^{-1}
=\bm{S}_t
&&
t=0,1
\label{st}
\end{align}
Regarding (\ref{genn}), (\ref{st}) is the multi-sector version of (\ref{pppn}).
In other words, networks transformation between $t=0,1$ is endogenized by general equilibrium feedback under the restoring productivity growths $\Delta \ln \hat{\bm{\tau}}_t$ and the sectoral technologies embodied in the restoring parameters $(\hat{\bm{\alpha}}, \hat{\bm{\gamma}})$ given by (\ref{alphai}, \ref{gammai}).
\hyperlink{app2}{Appendix 2} provides analyses of the CCES substitution elasticities among different factor inputs.

\subsection{Aggregate Fluctuations}
By the mapping $\mathcal{E}\left(\bm{\tau}; r,w \right) = \bm{p}$ under (\ref{emm}), we can empirically study the macroeconomic influences of microeconomic productivity shocks through Monte Carlo simulation.
Let us impose artificial sectoral productivity shocks in the form of iid geometric Brownian motions, i.e., $\ln \bm{\tau} \sim \mathcal{N}\left( 0, \sigma^2 \ell \right)$, where $\ell$ denotes the length of time over which the growth is measured.
For simplicity, we evaluate aggregate macroeconomic fluctuations 
by $- \left( \ln \bm{p} \right) \bm{1}^\intercal/J$, or the change in GDP growth in terms of a price index evaluated by the representative household's Cobb-Douglas utility parameters, which we set all equal ($1/J$), following \citet{aceECTA}.
In this section, we study not only CCES but also simple and Leontief economies relative to a Cobb-Douglas economy.
A simple economy refers to the case in which there are no sectoral interactions and where (\ref{emm}) is reduced as $\bm{p} = \bm{\tau}^{-1}$, such that macroeconomic fluctuations are evaluated by the equal-weighted average of sectoral productivity growth levels, i.e., $-\left( \ln \bm{p} \right) \bm{1}^\intercal/J = \left( \ln \bm{\tau} \right) \bm{1}^\intercal/J$.

Note that a CCES unit cost function reduces to a (multifactor) CES unit cost function if all elasticities are the same, i.e., $\gamma_{i}=\gamma$ (for all $j$ while we omit the index):
\begin{align}
q 
&= \tau^{-1} \left( \hat{\alpha}_I (p_I)^\gamma + (1-\hat{\alpha}_I) \left(  \hat{\alpha}_{I-1} (p_{I-1})^\gamma + (1-\hat{\alpha}_{I-1})\left( \hat{\alpha}_{I-2}(p_{I-2})^\gamma +\cdots\right) \right) \right)^{\frac{1}{\gamma}}
\\
&= \tau^{-1} \left( \hat{\alpha}_I (p_I)^\gamma + (1-\hat{\alpha}_I) \hat{\alpha}_{I-1} (p_{I-1})^\gamma 
+ (1-\hat{\alpha}_I) (1-\hat{\alpha}_{I-1}) \hat{\alpha}_{I-2} (p_{I-2})^\gamma 
\cdots \right)^{\frac{1}{\gamma}}
\end{align}
The $i$th share parameter is $(1-\hat{\alpha}_{I})(1-\hat{\alpha}_{I-1})\cdots(1-\hat{\alpha}_{i+1})\hat{\alpha}_{i}$, which we know from (\ref{alphai}) equals $s_{i1}$ in the reference period, as we normalize the model at $t=1$.\footnote{Normalization refers to setting all prices of the reference period at unity.  In this case, we set $p_{11}=p_{21}=\cdots=p_{I1}=r_{1}=w_{1}=1$, and thus, $\hat{\pi}_{11}=\hat{\pi}_{21}=\cdots=\hat{\pi}_{I1}=1$, in all sectoral productions.}
In what follows, we write the reference production networks as $\mathbf{S}_1=\mathbf{A}$, and correspondingly, 
$
\bm{1}-\sum_{i=1}^I \bm{s}_{i1}
= \bm{a}_{0}
$. 
Note further that primary factor prices must be set in the reference period i.e., $r=w=1$.
We can then rewrite the above identity as follows:
\begin{align}
\left( q \tau \right)^\gamma
=
\sum_{i=1}^I s_{i1} (p_{i})^\gamma
+s_{K1} r^\gamma + s_{L1} w^\gamma 
=
\sum_{i=1}^I a_{i} (p_{i})^\gamma +a_{0} 
\label{mces}
\end{align}
For a Cobb-Douglas economy ($1-\gamma = 1$), (\ref{mces}) can be reduced, using l'H{\^o}spital's rule, as follows:
\begin{align}
\ln q \tau 
= \lim_{\gamma \to 0} \frac{\ln \left(\sum_{i=1}^I a_{i} (p_i)^\gamma + a_0 \right)}{\gamma}
= \lim_{\gamma \to 0} \frac{\sum_{i=1}^I a_{i} (p_i)^\gamma \ln p_i}{\sum_{i=1}^I a_i (p_{i})^\gamma + a_0 }
= {\sum_{i=1}^I a_{i} \ln p_i }
\end{align}
For a Leontief economy ($1-\gamma = 0$), (\ref{mces}) reduces as follows:
\begin{align}
q \tau = \sum_{i=1}^I a_{i} p_i +a_0
\end{align}
Hence, as prices equilibrate ($\bm{p}=\bm{q}$), the two economies have closed-form equilibrium solutions. 
\begin{align}
\ln \bm{p} &= -(\ln \bm{\tau} )\left[ \mathbf{I} - \mathbf{A}  \right]^{-1}
&&
\text{Cobb-Douglas} \label{CD}
\\
\bm{p} &= \bm{a}_0 \left[ \left< \bm{\tau} \right> - \mathbf{A} \right]^{-1}
&&
\text{Leontief} \label{LT}
\end{align}

We impose the same artificial productivity growth shocks $\ln \tilde{\bm{\tau}}$, where $\ln \tilde{\tau}_{j}=( \ln \tau_{j}{(1)}, \cdots, \ln \tau_{j}{(D)} )$ is a string of $D=300$ draws from a normal distribution $\mathcal{N}\left( 0, \sigma^2 \ell \right)$, into the mapping $\bm{p} = \mathcal{E}\left( \bm{\tau}; 1,1 \right)$ with different alternative underlying economies, namely, Cobb-Douglas, Leontief, simple and restoring CCES.
Correspondingly, let ${\bm{\tau}}(d) = \left( \tau_1(d), \cdots, \tau_J (d) \right) $ denote the $d$th (sector-wide productivity) shock, where the shocks are indexed by $d = 1, \cdots, D$.
For our purpose, we use volatility $\sigma$ that amounts to $10$\% per year or $1068$ ppm per hour.
The procedure applied here is to plug ${\bm{\tau}}(d)$ into (\ref{CD}) for Cobb-Douglas, (\ref{LT}) for Leontief, and (\ref{emm}) for restoring CCES economies to calculate the corresponding equilibrium price $\bm{p} (d)$ and evaluate the (simulated) aggregate fluctuations, i.e., $-(\ln \bm{p}(d)) \bm{1}^\intercal /J$ for all $d=1, \cdots, D$.
In this way, the differences in the simulated aggregate fluctuations can be attributed to the differences in the underlying set of alternative economies for each $d$.
\begin{figure}[t!]
\centering
    \includegraphics[width=0.495\textwidth]{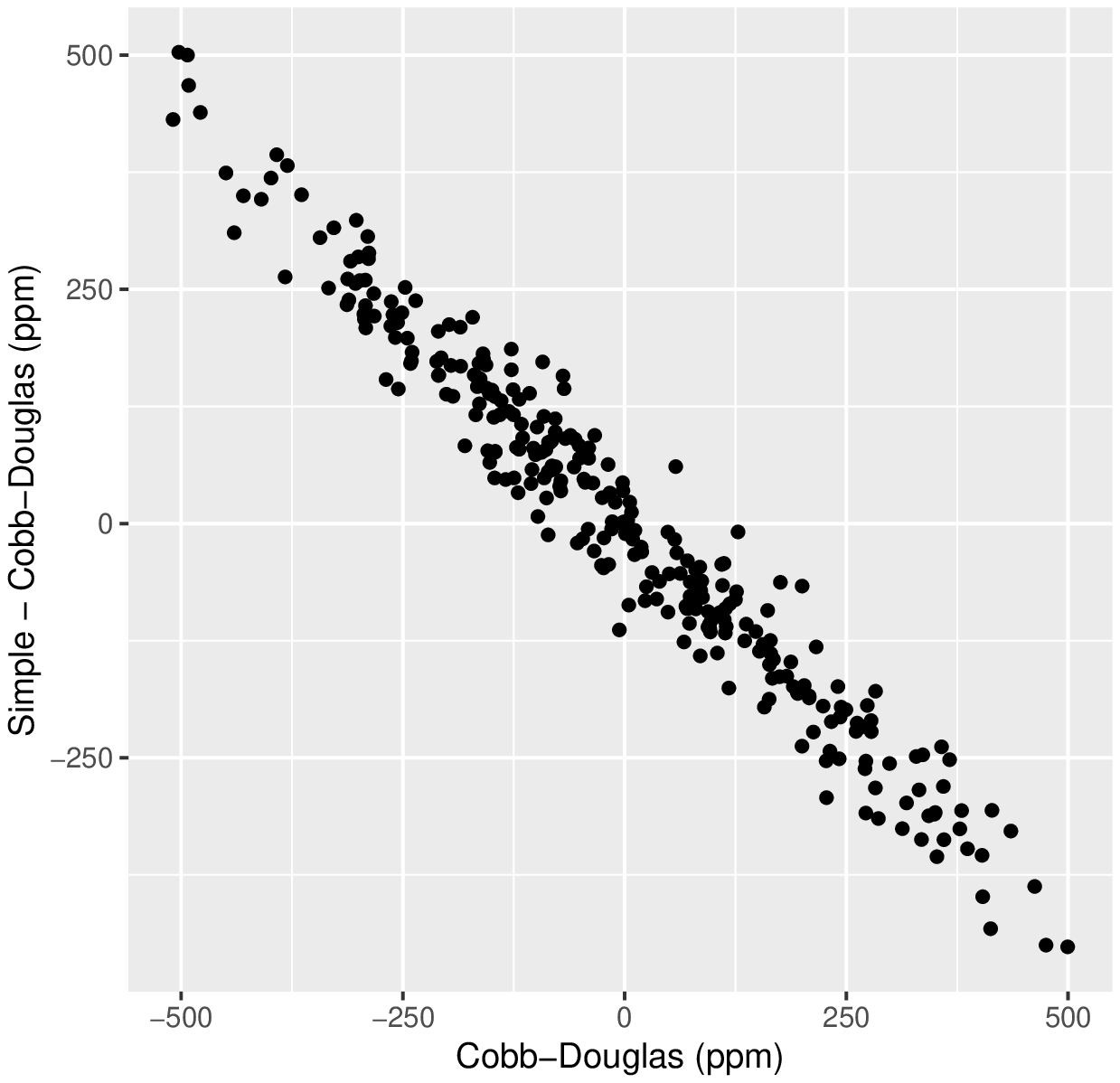}
    \includegraphics[width=0.495\textwidth]{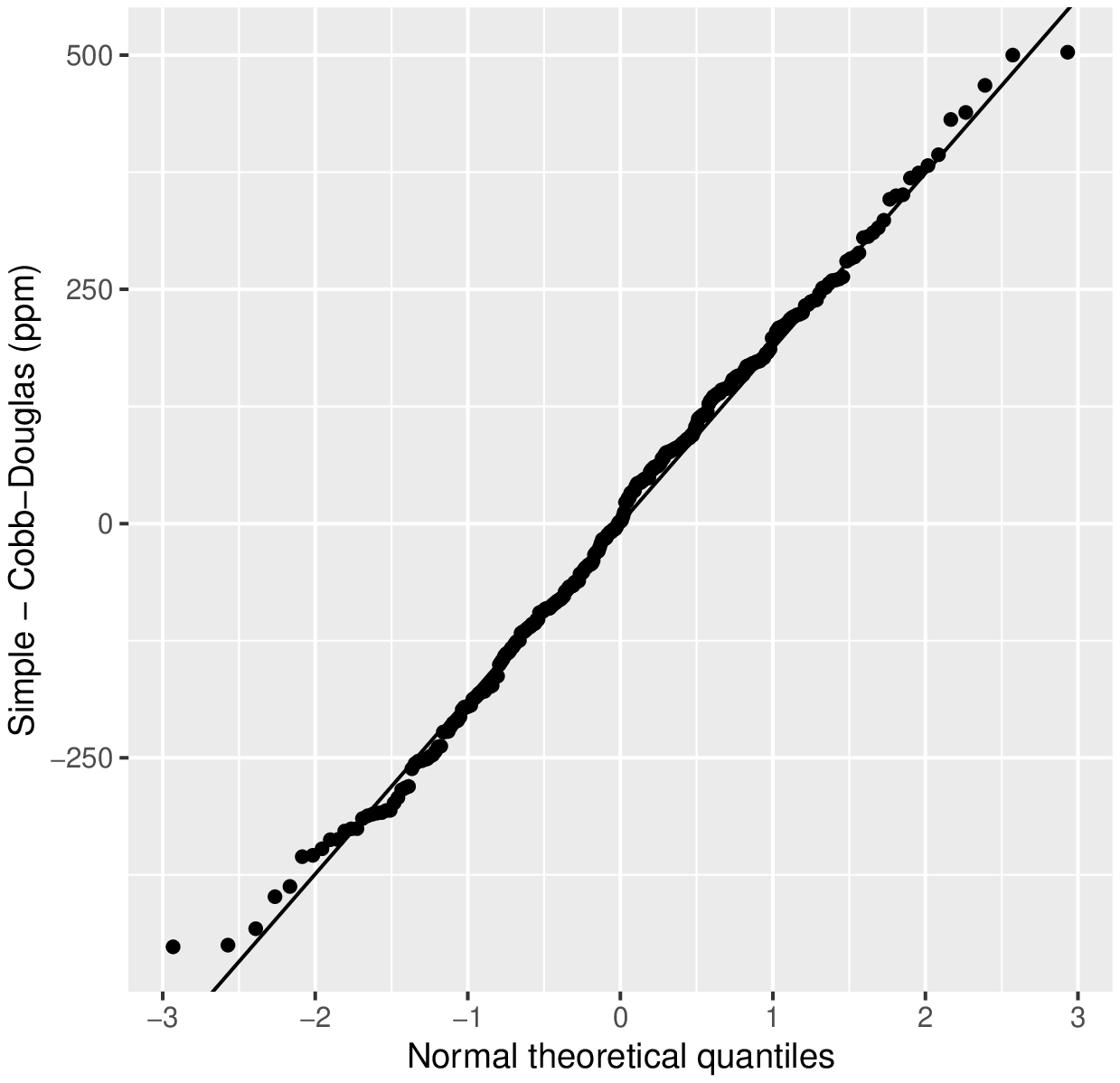}
    \caption{
Left: Differences in simulated aggregate fluctuations between simple and Cobb-Douglas economies against the simulated aggregate fluctuations in the Cobb-Douglas economy.
Right: QQ-plot of the distribution of the vertical axis variable indicating its normality.
The volatility of artificial productivity growth shocks is 10\% per year or 1068 ppm per hour.
} \label{SMCD}
\end{figure}

\begin{figure}[t!]
\centering
    \includegraphics[width=0.495\textwidth]{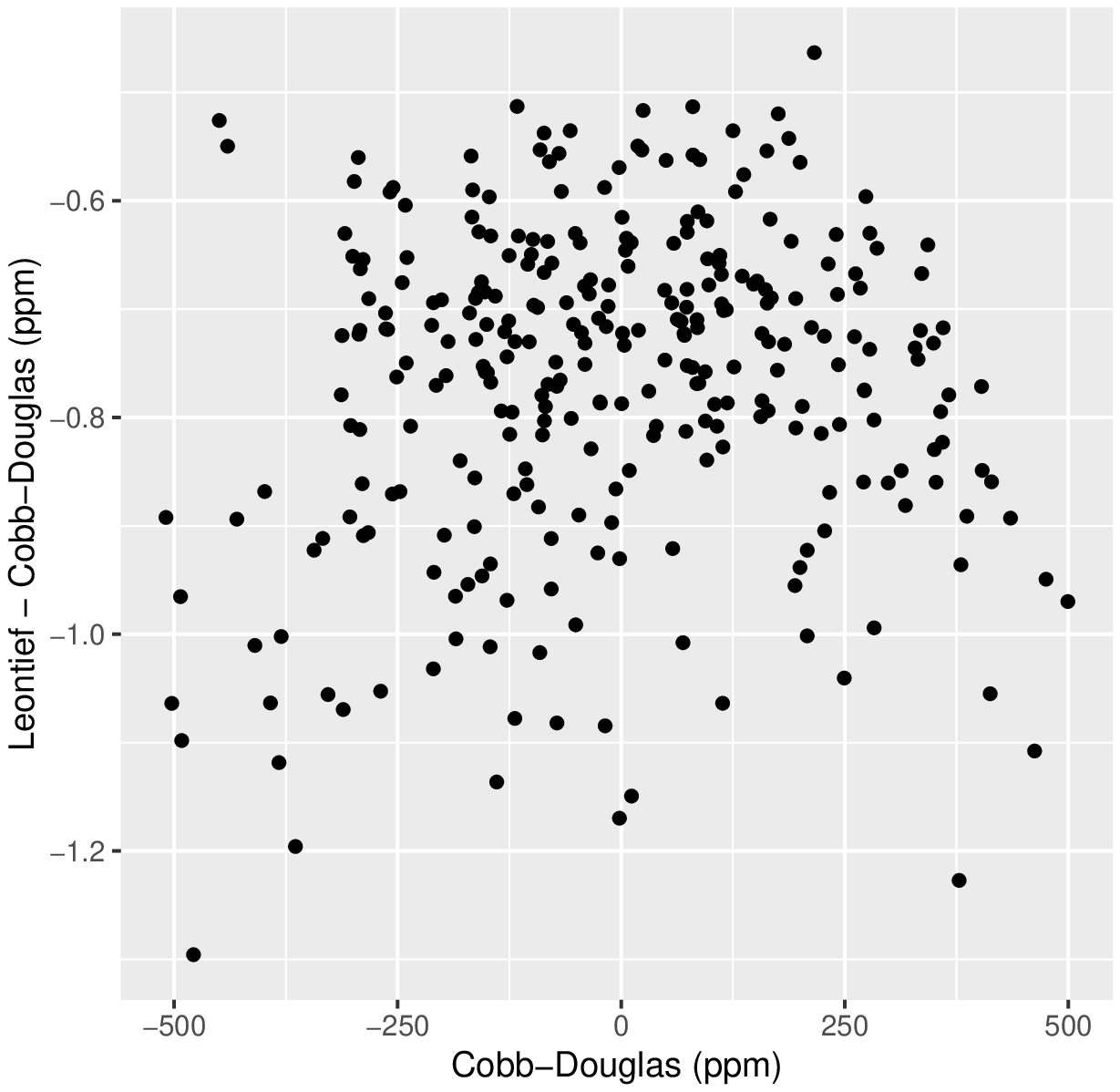}
    \includegraphics[width=0.495\textwidth]{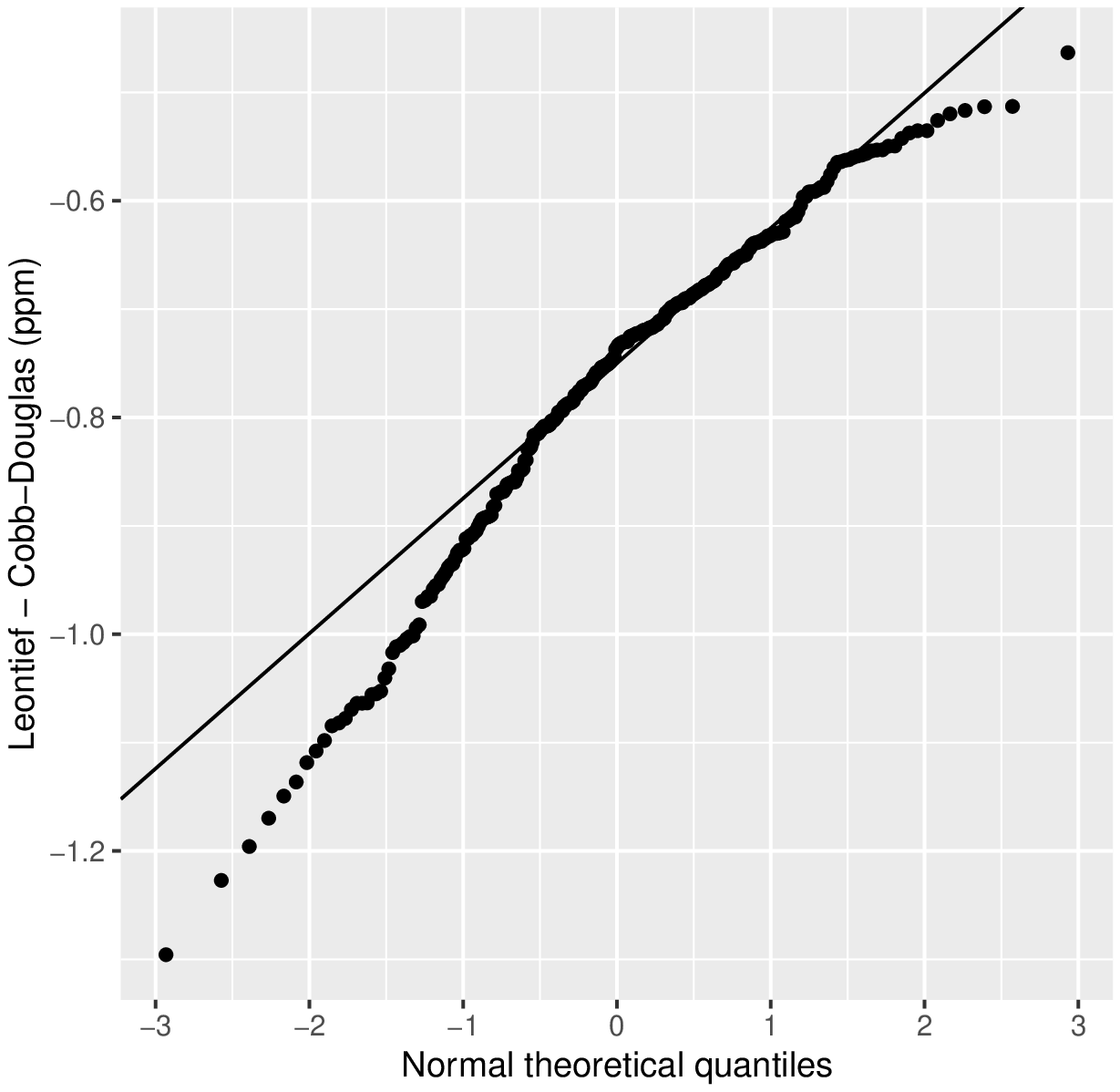}
    \caption{
Left: Differences of simulated aggregate fluctuations between Leontief and Cobb-Douglas economies against the simulated aggregate fluctuations of Cobb-Douglas economy.
Right: QQ-plot of the distribution of vertical-axis variable indicating its non-normality.
The volatility of artificial productivity growth shocks is 10\% per year or 1068ppm per hour.
}
    \label{syn2}
\end{figure}

In parallel to previous studies, we first compare the differences between simple and Cobb-Douglas economies (see Figure \ref{SMCD}).
While the growth shocks are aggregated using equal weights in the case of the simple economy, the aggregate volatility is larger if the growth shocks are aggregated by with unequal weights.
\citet{gabaixECTA} showed that aggregate volatility derived with equal weights, i.e., ${\sigma}/{\sqrt{J}}$, becomes ${\sigma}/{\ln J}$ if the weights are granular (i.e., distributed exponentially).
\citet{aceECTA} found that a similar volatility boost is possible for the Cobb-Douglas economy, where the growth shocks are aggregated by the Leontief inverse (\ref{CD}).\footnote{The aggregate volatility for the granular economy must be $\sqrt{J}/\ln J = 3.3$ times larger regarding the dimension of our models ($J=385$).
Our simulated aggregate volatility for the Cobb-Douglas economy (233 ppm per hour or 2.2\% per year) based on input-output table for Japan is 4.3 times larger than that for the simple economy (i.e., $1068/\sqrt{385}=54$ ppm per hour).}
The aggregate fluctuations are evaluated by $\left( \ln \bm{\tau}(d)\right) \bm{1}^\intercal/J$ for the simple economy and by $\left( \ln \bm{\tau}(d) \right) \left[ \mathbf{I} - \mathbf{A}  \right]^{-1} \bm{1}^\intercal/J$ for the Cobb-Douglas economy.
As the artificial productivity growth shocks, which are normally distributed, are linearly aggregated in both cases, the aggregate fluctuations must also be normally distributed in both cases.\footnote{\citet{aceAER} show that a Domar-weighted aggregation (i.e., linear aggregation under Cobb-Douglas economy) of heavy-tailed shocks can produce heavy-tailed aggregate fluctuatoins.  In contrast, Figures \ref{syn2} and \ref{syn3} show that heavy-tailed aggregate fluctuations can be generated by normally distributed shocks entering nonlinear (Leontief and CCES) economy.}
This can be verified in Figure \ref{SMCD} (right).

In the case of the Leontief economy, it is obvious from (\ref{LT}) that the aggregated fluctuations $-(\ln \bm{p}) \bm{1}^\intercal /J$ are \text{nonlinear} with respect to the productivity growth shocks $\ln \bm{\tau}$.
This nonlinearity makes the simulated aggregate fluctuations for the Leontief economy depart from a normal distribution (see Figure \ref{syn2}).
Note that the simulated aggregate fluctuations for the two economies are very similar, but the Leontief economy always provides less welfare (GDP growth) than in the Cobb-Douglas case.
This is because, under the same factor prices, the unit cost of Cobb-Douglas technology is always less than that of the Leontief because Cobb-Douglas has more alternative technologies in addition to the one it shares with Leontief.
The simulations are, however, limited to a short span of time (an hour) with a standard deviation of 1068 ppm due to our computational capacity in processing the convergence iterations (\ref{emm}) for the restoring CCES economy, which might entail complex nonlinearities (noconcavities) in completely restoring the two states.

Although we could have simulated a longer period for the Leontief economy, for which the equilibrium calculation only involves matrix inversion, we opted to maintain comparability with the restoring CCES economy to potentially reveal more significant differences from the Cobb-Douglas case.
It is clear from (the vertical axes of) Figures \ref{syn2} and \ref{syn3} that the restoring CCES economy reveals greater differences from Cobb-Douglas than does Leontief, with the tendency for the differences to be negative.
These figures imply that restoring CCES underperforms Leontief in terms of increases in GDP growth from fair (i.e., zero mean) productivity shocks.
Note that while restoring CCES entirely replicates the two observed states, this does 
not necessarily mean that the aggregator functions are all concave and that potential technology substitution will always be cost improving.\footnote{We could have introduced concavity constraints in addition to the first-order conditions in solving NLP (\ref{nlp}) to estimate the CCES parameters and spoiled the restoring property; however, we opted to pursue the opposite.}
However, the restoring CCES can occasionally exhibit greater GDP growth than Cobb-Douglas, and interestingly, we observe that distribution departs from normality in both directions.
For reference, Table \ref{tab_1} summarizes the statistical moments of the simulated aggregate fluctuations displayed in Figures \ref{SMCD}, \ref{syn2} and \ref{syn3}.
\newcolumntype{.}{D{.}{.}{4}}			
\begin{figure}[t!]
\centering
    \includegraphics[width=0.495\textwidth]{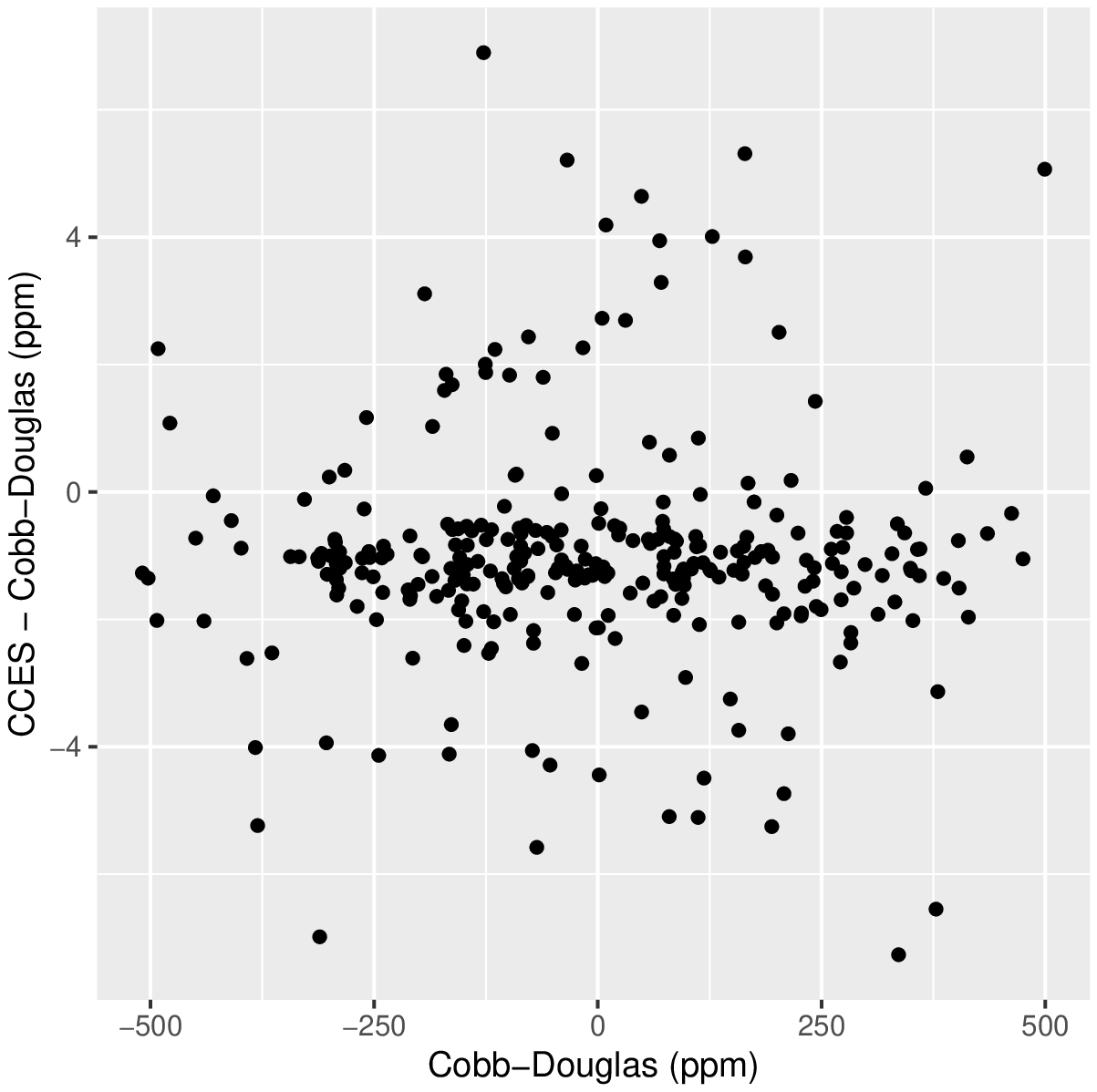}
    \includegraphics[width=0.495\textwidth]{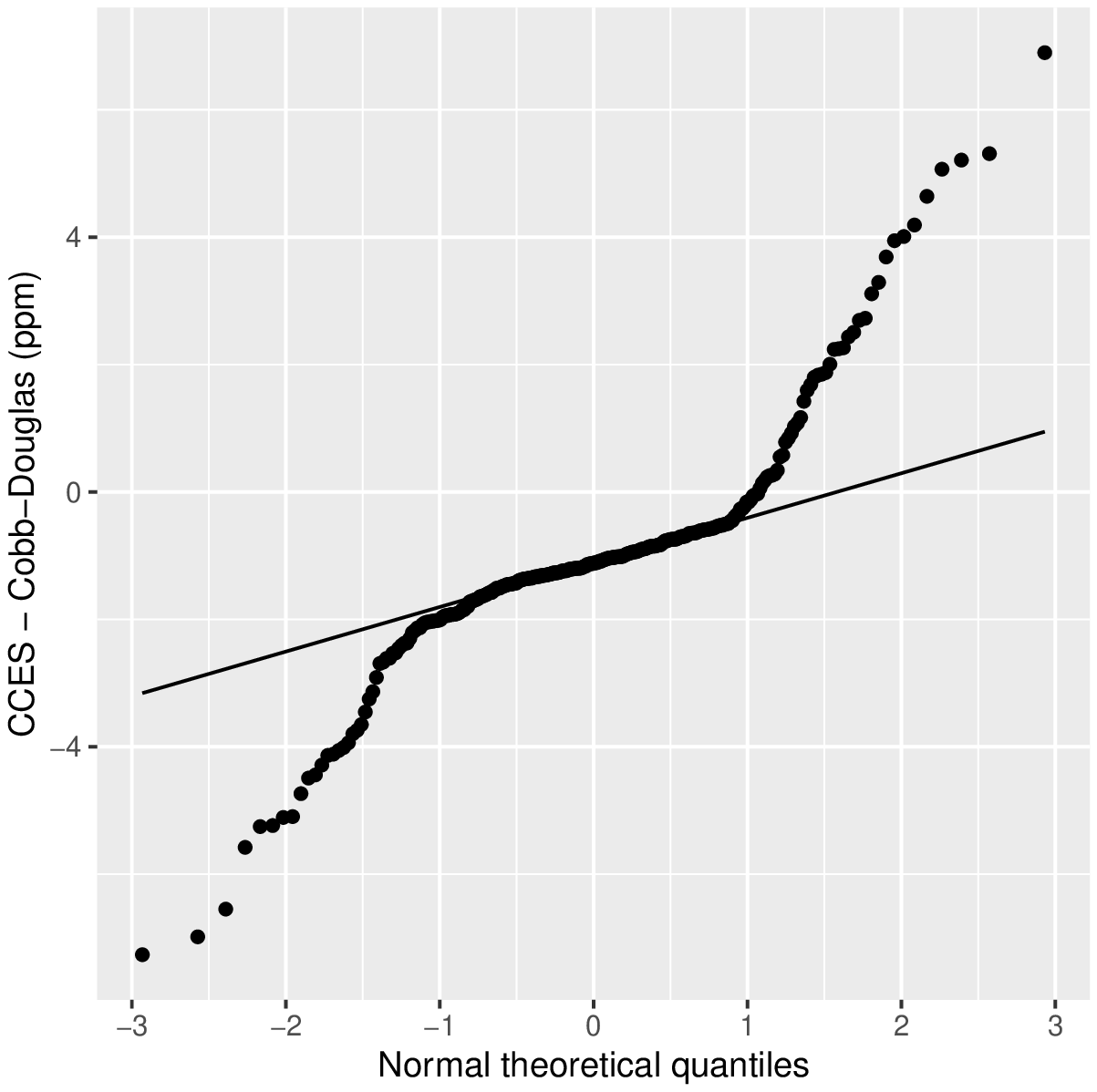}
    \caption{
Left: Differences in the simulated aggregate fluctuations between restoring CCES and Cobb-Douglas economies against the simulated aggregate fluctuations of the Cobb-Douglas economy.
Right: QQ-plot of the distribution of the vertical axis variable indicating its non-normality.
The volatility of artificial productivity growth shocks is 10\% per year or 1068 ppm per hour.}
    \label{syn3}
\end{figure}
\begin{table}[t!]						
\caption{Statistical moments of simulated aggregate fluctuations corresponding to the vertical values of Figures \ref{SMCD}, \ref{syn2} and \ref{syn3}.} \label{tab_1}\vspace{-10pt}			
\begin{center}						
\begin{tabular}{l...}	
\hline\noalign{\smallskip}			
	&	\multicolumn{1}{c}{Simple $-$ Cobb-Douglas}	&	\multicolumn{1}{c}{Leontief $-$ Cobb-Douglas}	&	\multicolumn{1}{c}{CCES $-$ Cobb-Douglas}	\\
\hline\noalign{\smallskip}			
Mean	&	12.552	&	-0.768	&	-1.006	\\
Standard Deviation	&	244.697	&	0.149	&	1.783	\\
Skewness	&	-0.076	&	-0.745	&	0.642	\\
Excess Kurtosis	&	-0.545	&	0.393	&	4.077	\\
\hline			
\end{tabular}							
\end{center}							
\end{table}

%\clearpage
\section{Dynamic General Equilibrium}
\subsection{Representative Household}
We consider a representative household, the utility of which is modeled as a multifactor CES aggregator function as follows:
\begin{align}
u\left( \bm{h} \right)
=\left( (\mu_{1})^{\frac{1}{1-\lambda}} (h_{1})^{\frac{\lambda}{1-\lambda}} + \cdots + (\mu_{I})^{\frac{1}{1-\lambda}} (h_{I})^{\frac{\lambda}{1-\lambda}} \right)^{\frac{\lambda-1}{\lambda}}
\end{align}
where $\mu_i$ and $1-\lambda$ denote the $i$th share parameter and the elasticity of substitution, respectively.
Given budget ${H}$ and price of all goods $\bm{p}=\left( p_1, \cdots, p_I \right)$, the household determines the consumption schedule $\bm{h}=\left( h_1, \cdots, h_I \right)$ that maximizes its utility, where ${H}=p_1 h_i + \cdots + p_I h_I$ must hold.
After some calculations, we arrive at the following multifactor CES indirect utility function:
\begin{align}
v\left( \bm{p}; {H} \right)
= {H} \left( \mu_i (p_i)^\lambda + \cdots + \mu_I (p_I)^\lambda
 \right)^{-{1}/{\lambda}}
 ={H} / \psi ( \bm{p})
 \label{vpsi}
\end{align}
We define the price index $\psi$ as above for later convenience.

By applying Roy's identity, i.e., $h_i = - \frac{\partial v}{\partial p_i}/\frac{\partial v}{\partial {H}}$, we have the following expansion for the expenditure share of the $i$th good, denoted by $b_i$:
\begin{align}
b_i = \frac{p_i h_i}{p_1 h_1 + \cdots + p_I h_I} = \frac{\mu_i (p_i)^\lambda}{\mu_1 (p_1)^\lambda + \cdots + \mu_I (p_I)^\lambda} = \mu_i \left( \frac{p_i}{\psi (\bm{p})} \right)^{\lambda}
\label{bemu}
\end{align}
Now, we know that parameter $\lambda$ can be estimated by the variety of expenditure shares and prices. 
By taking logs and indexing samples by $t$, we have the following expansion with the error term $\epsilon_{it}$: 
\begin{align}
\ln b_{it} = \ln \mu_i - \lambda \ln \psi_t + \lambda \ln p_{it} + \epsilon_{it}
\end{align}
The parameter $\lambda$ can thus be estimated by fixed effect regression i.e., 
\begin{align}
\Delta \ln b_{it} = - \lambda \Delta \ln \psi_t + \lambda \Delta \ln p_{it} + \Delta \epsilon_{it}
\label{fe}
\end{align}
In what follows, we will use item-wise observations for $i=1,\cdots,I$ with a minimum periodical dimension of two ($t=0,1$).

In the estimation of (\ref{fe}), at least two issues must be addressed.
The first is the endogeneity of the regressor, and the second is the heteroskedasticity of the error term.
Regarding endogeneity (i.e., endogeneity due to the anticipated reverse causality that a representative household's expenditures can affect commodity prices), we perform instrumental variables estimation using sector-wise restoring productivity growths with CCES, i.e., the $\Delta \ln \hat{\tau}_{it}$ previously measured (see Figure \ref{fig_tfpg}).
Sectoral productivity growths and the error terms (representative consumer's taste shocks) must be uncorrelated.
Regarding heteroskedasticity, we consider potential measurement errors for log-difference transformations of two stochastic variables $(b_{i0}, b_{i1})$.
We assume that these are normally distributed random variables with mean $(b_{i0}, b_{i1})$ and some homoskedastic variance $(\sigma_b)^2$.
In this case, the dependent variable's variance can be approximated as follows:
\begin{align}
\Var \left( \ln b_{i1} - \ln b_{i0} \right) \approx (\sigma_b)^2 \left( \frac{1}{(b_{i1})^2} + \frac{1}{(b_{i0})^2} \right) = (\sigma_b)^2 (\nu_i)^2
\end{align}

Below, we display the result of a weighted two-stage least squares estimation using $\nu_i$ (defined above) as weights and both $\Delta \ln \hat{\tau}_{it}$ and $e^{\Delta \ln \hat{\tau}_{it}}$ as instruments.
Standard errors are shown in parentheses:
\begin{align}
\ln b_{i1}/b_{i0} = \underset{(0.00850)}{0.00561} + \underset{(0.35218)}{1.09631} \ln p_{i1}/p_{i0} 
\label{result}
\end{align}
Considering the first-stage F statistic ($\text{F}(2, 265)= 119.57$), we are not concerned about a weak instrument problem. 
In testing the instruments for overidentifying restrictions (Sargan $\chi^2(1) =0.2917$, Basmann $\chi^2(1) = 0.2887$), we do not reject the null hypothesis that at least one of the instruments is endogenous.
Regarding the Durbin and Wu--Hausman test for regressor endogeneity (Durbin $\chi^2 (1) = 10.5032$, Wu-Hausman F(1, 265) $=10.8093$), we reject the null hypothesis that the regressor is exogenous.
In what follows, we therefore use $\hat{\lambda} = 1.096$.
Moreover, it must be appropriate to use $\hat{\mu}_i=b_{i1}$ because we standardize the model at $t=1$, where according to (\ref{bemu}), $\bm{p}_{1} = \bm{1}$ leads to $b_{i1}=\mu_{i}$ in light of the assumption that $\sum_{i=1}^I \mu_{i}=1$.
Hence, the empirical price index function $\psi$ can be specified as follows:
\begin{align}
\psi (\bm{p}) = \left( b_{11}(p_{1})^{\hat{\lambda}} + \cdots + b_{I1}(p_{I})^{\hat{\lambda}} \right)^{1/\hat{\lambda}}
\label{empindex}
\end{align}

\subsection{Social Benefit Assessment}
To perform the assessment, we introduce an infinitely lived, unique representative household, the utility of which is modeled by a multifactor CES aggregator as follows:\footnote{In what follows $t=2,3\cdots$ designates a period in the future and not a sampled period in the past.}
\begin{align}
\sum_{t=0}^\infty \beta^t u\left( \bm{h}_t \right)
=\sum_{t=0}^\infty \beta^t \left( (\mu_{1})^{\frac{1}{1-\lambda}} (h_{1t})^{\frac{\lambda}{1-\lambda}} + \cdots + (\mu_{I})^{\frac{1}{1-\lambda}} (h_{It})^{\frac{\lambda}{1-\lambda}} \right)^{\frac{\lambda-1}{\lambda}}
\end{align}
where $\beta$ is the discount factor.
The share parameter is denoted by $\mu_i$ for $i=1,\cdots,I$ where $\sum_{i=1}^I \mu_i = 1$, and the elasticity of substitution is denoted by $\lambda$.
These parameters are to be replaced by the estimates given in the previous section.
The representative household maximizes the above objective function subject to the following economy-wide budget constraint:
\begin{align}
{H}_{t} + z_t \rho \left( K_{t+1} - (1-\delta) K_{t} \right) + M_t
=r_t K_t + w_t L_t
\label{bc}
\end{align}
where, ${H} = \sum_{i=1}^I p_i h_i$ (the household's budget), $K = \sum_{j=1}^J K_{j}$ (total capital service), $L = \sum_{j=1}^J L_{j}$ (total labor), and $M = \sum_{i=1}^I p_i m_i$.
The second term on the left-hand side corresponds to fixed capital formation $G=\sum_{i=1}^I p_{i} g_{i}$.
Note that the above balance is equivalent to $\sum_{i=1}^I p_i f_i = \sum_{j=1}^J e_j$, where $f_i = h_i + g_i + m_i$, regarding the input-output tables (\ref{data}).
We denote the ratio between the capital stock and capital service by $\rho$ and use $z$ to denote the price of capital. 
The depreciation rate is denoted by $\delta$.
The first-order condition of the representative household's problem yields the following Euler equation:
\begin{align}
\beta \frac{z_{t+1}\rho(1-\delta)+r_{t+1}}{z_t \rho}
=
\frac{
\frac{\partial u}{\partial h_{it}} \frac{1}{p_{it}}
}
{
\frac{\partial u}{\partial h_{i t+1}} \frac{1}{p_{i t+1}}
}
=\frac{\psi(\bm{p}_{t+1})}{\psi(\bm{p}_t)}
\label{euler}
\end{align}
where we use marginal utility of money, i.e., $\frac{\partial u}{\partial h_i} \frac{1}{p_i} = \frac{\partial v}{\partial {H}}$ and (\ref{vpsi}) to derive the second identity.

In what follows, we show how we retrieve physical quantities to evaluate the general equilibrium in terms of economic welfare.
Below is the breakdown of the budget constraints (\ref{bc}) for $t=0,1$:
\begin{align}
{H}_{0} + G_0 + M_0
=r_0 K_0 + w_0 L_0  
&&
G_{0} = z_{0} \rho \left( K_{1} - (1-\delta) K_{0} \right)
\label{bu0}
\\
{H}_{1} + G_{1} + M_1
=r_1 K_1 + w_1 L_1 
&&
G_{1}=z_{1} \rho \left( K_{2} - (1-\delta) K_{1} \right) 
\label{bu1}
\end{align}
where the terms in the equations on the left side are all available from the two-period linked input-output tables. 
Thus, we know from (\ref{bu0}) and (\ref{bu1}) that $z_0 \rho = \frac{G_0}{K_1 - (1-\delta) K_0}$, and we can borrow $\delta$ from external sources.\footnote{We use a five-year value, $\delta=1-(1-0.125)^5$, following \citet{nomurasuga}.}
On the other hand, we do not observe $K_2$, so we use the Euler equation (\ref{euler}) for the two periods described below to measure $z_1\rho$:
\begin{align}
\beta \frac{z_{1}\rho(1-\delta)+r_{1}}{z_0 \rho}
=\frac{\psi(\bm{p}_{1})}{\psi(\bm{p}_0)}
\label{euler01}
\end{align}
where we can use the empirical price index function (\ref{empindex}) for $\psi$ and external sources for $\beta$.\footnote{We use a five-year value $\beta=(1+0.03)^{-5}$, following \citet{kawasaki, idagoto}.} 
We can then use (\ref{bu1}) to determine that $K_2 = \frac{G_1}{z_1\rho}+(1-\delta)K_1$.
Finally, the price elasticity of fixed capital formation $\eta_K$ can be measured as follows:
\begin{align}
\eta_K = \frac{\left({K_2 - (1-\delta)K_1}\right) - \left({K_1-(1-\delta)K_0}\right)}{z_1\rho - z_0\rho}
\frac{z_0\rho}{K_1-(1-\delta)K_0} 
= -0.80
\end{align}
We will use this elasticity to link price with quantity and hence the welfare of the economy.

Below, we consider whether, at the reference point $t=1$, the productivity is different from $\hat{\bm{\tau}}_1 = \bm{1}$ and evaluate the potential difference in the welfare of the economy. 
We denote by $\check{\bm{\tau}}_1$ the alternative productivity at $t=1$ and indicate all variables under this productivity by a check.
An alternative equilibrium price $\check{\bm{p}}_1$ can be obtained by mapping $\mathcal{E}\left( \check{\bm{\tau}}_1; r_1, w_1 \right)$ under restoring CCES.
Quantitative differences will be evaluated for $K$, $L$, and $h_i$, among others, while we hold $r$, $w$, and $M$ fixed for sake of simplicity.\footnote{Specifically, we assume a unit price elasticity of $m_i$ for all $i$, where $M=\sum_{i}^I p_i m_i$ is invariant to price changes.}
First, we use the following modification of (\ref{euler}) to evaluate $\check{z}_1\rho$ from $\check{\bm{p}}_1$.
\begin{align}
\frac{\check{z}_1\rho (1-\delta) + r_1}{{z}_1\rho (1-\delta) + r_1}
=
\frac{\psi(\check{\bm{p}}_1)}{\psi(\bm{p}_1)}
\end{align}
Then, we evaluate $\check{K}_2$ or $\check{G}_1 = \check{z}_1\rho(\check{K}_2 - (1-\delta)K_1)$ by the elasticity $\eta_K$, i.e.,
\begin{align}
\eta_K = \frac{({\check{K}_2 - (1-\delta)K_1}) - ({K_1-(1-\delta)K_0})}{\check{z}_1\rho - z_0\rho}
\frac{z_0\rho}{K_1-(1-\delta)K_0} 
\end{align}
We will decompose $\check{G}_1$ into items by constant ratios.
Each component of $M_1$ is assumed to be constant.
Regarding $\check{H}_1$, which will be given recursively by (\ref{feed}), we decompose it into items according to (\ref{bemu}):
\begin{align}
\check{p}_{i1}\check{h}_{i1} = \mu_i \left( \frac{\check{p}_{i1}}{\psi (\check{\bm{p}}_1)} \right)^{{\lambda}} \check{H}_1
&&
\check{p}_{i1}\check{g}_{i1} = \kappa_i  \check{G}_1
&&
\check{p}_{i1}\check{m}_{i1} = {p}_{i1}{m}_{i1}
\label{back}
\end{align}
Here, $\kappa_i = p_{i1}g_{i1}/G_{1}$ is assumed constant, and we use $\mu_i=b_i$ and $\lambda = \hat{\lambda}$, in our empirical modeling (\ref{empindex}). 
We use these quantities $(\check{\bm{h}}_1, \check{\bm{g}}_1, \check{\bm{m}}_1)$ of final demand to evaluate alternative labor $\check{L}_1$ through input-output analysis under the alternative equilibrium cost-share structure (input coefficient matrix) extrapolated under restoring CCES.
The procedure can be described as follows:
\begin{align}
\check{L}_1 = \check{\bm{a}}_{L} \bigl[ \mathbf{I} - \check{\mathbf{A}} \bigr]^{-1}
\langle \check{\bm{p}}_1 \rangle \bigl[ \langle \check{\bm{h}}_1 \rangle + \left< \check{\bm{g}}_1 \right> + \left< \check{\bm{m}}_1 \right>  \bigr]
\end{align}
where labor intensity and input-output coefficients of the alternative equilibrium, denoted $\check{\bm{a}}_L$ and $\check{\mathbf{A}}$, respectively, are obtained by the following formula under restoring CCES:
\begin{align}
\langle \check{\bm{p}}_1, r_1, w_1 \rangle \nabla \bm{C} \left( \check{\bm{p}}_1, r_1, w_1  \right)
\langle \check{\bm{\tau}}_1 \rangle^{-1} \langle \check{\bm{p}}_1 \rangle^{-1}
= \bigl[ \check{\mathbf{A}}, \check{\bm{a}}_K, \check{\bm{a}}_L \bigr]^\intercal
\end{align}
Finally, $\check{H}_{1}$ is evaluated by the following alternative budget constraint and fed back into (\ref{back}) to eventually reach a solution for the alternative equilibrium.
\begin{align}
\check{H}_{1} + \check{z}_1 \rho \bigl( \check{K}_{2} - (1-\delta) K_{1} \bigr) + M_1
= r_1 K_1 +  w_1 \check{L}_1
\label{feed}
\end{align}
Social benefits and costs under the alternative productivity $\check{\bm{\tau}}_1$ are hence evaluated by the differences in representative household's (indirect) utility and labor provided, viz.,
\begin{align}
\text{Benefit}(\check{\bm{\tau}}_1)=\check{H}_1 \psi({\bm{p}}_1)/\psi(\check{\bm{p}}_1) - {H}_1
&&
\text{Cost}(\check{\bm{\tau}}_1)=w_1 \check{L}_1 - w_1 L_1
\label{bandc}
\end{align}

\subsection{Imposing Standard Productivity}
Here, we virtually impose the same level of productivity in each sector and assess how much welfare can be gained through its dynamic general equilibrium propagation.
The productivity increment for the $j$th sector is standardized according to the sector's magnitude of output at the reference point, $p_{j1} y_{j1}$.
Specifically, we define the standardized (or unit) productivity triggers as follows:
\begin{align}
\check{\bm{\tau}}_1(j)=\left( \tau_{11}, \cdots, \check{\tau}_{j1}, \cdots, \tau_{J1}  \right)
&&
\check{\tau}_{j1} = 1+\frac{\theta}{p_{j1} y_{j1}}
\end{align}
where $\theta$ denotes the standard increment of productivity in monetary value, common to all sectors, which we set arbitrarily to $\theta =1$ billion yen.
That is, all standardized productivities have a common unit of magnitude $\theta$.
We emphasize that $\theta$ is an indicator of magnitude and not the cost of imposing unit productivity in a sector's production.
This imposition of standardized productivity is assessed, with respect to its return in terms of net social benefit, by the following figure (as social return on productivity, SROP):
\begin{align}
\text{SROP}\left(\check{\bm{\tau}}_1(j)\right) = \frac{\text{Benefit}\left(\check{\bm{\tau}}_1(j)\right) - \text{Cost}\left(\check{\bm{\tau}}_1(j)\right)}{\theta}
\label{srop}
\end{align}
where the measure of welfare is normalized by $\theta$.
Note that SROP is relative and should not be interpreted as a return on investment because $\theta$ is not the cost but an indicator of magnitude.\footnote{In that sense, we would not have needed to normalize the net benefit by $\theta$, but we leave it as it is.}

\begin{figure}[t!]
\centering
    \includegraphics[width=0.495\textwidth]{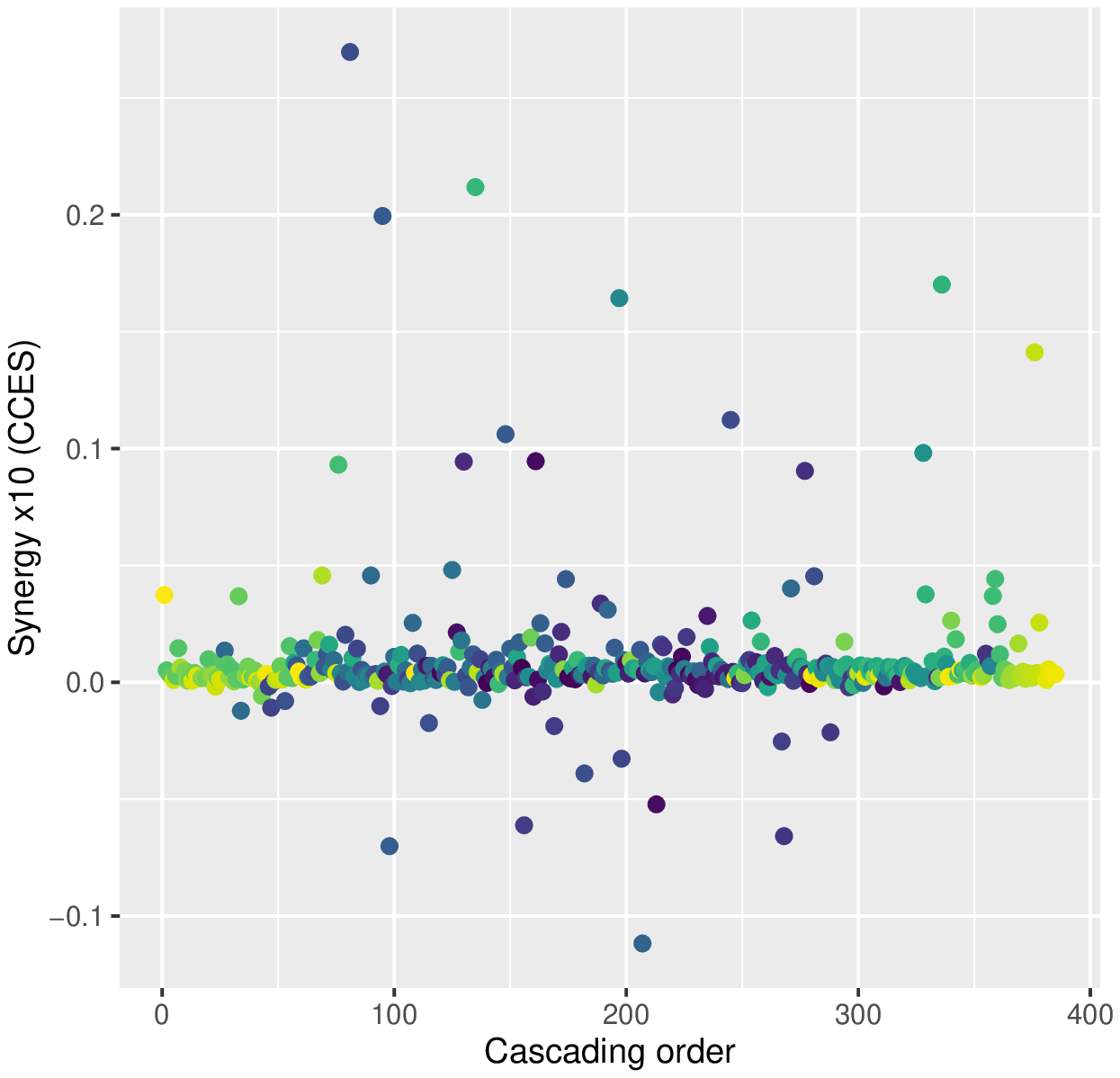}
    \includegraphics[width=0.495\textwidth]{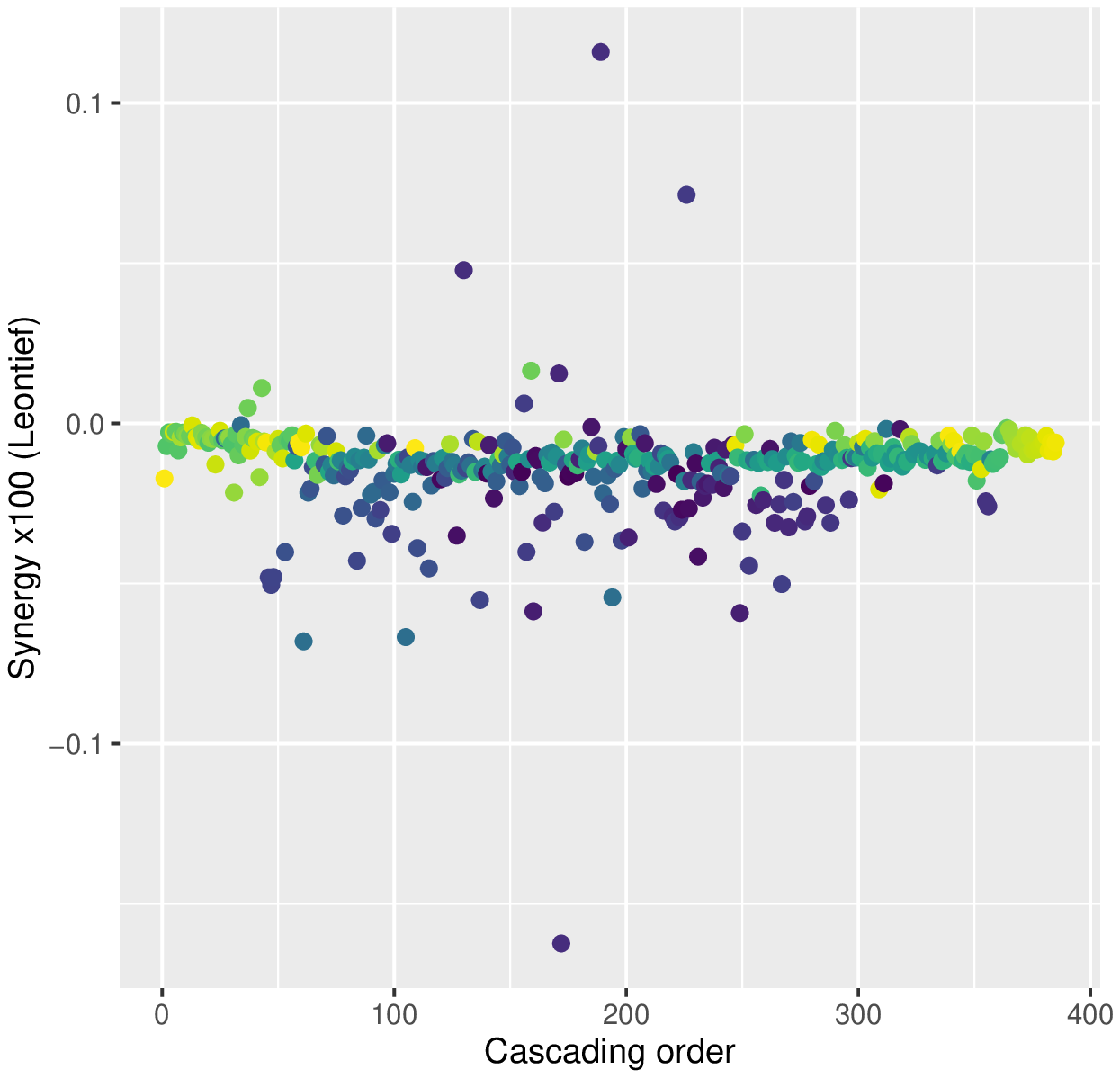}
    \caption{
Left: Synergy (defined as (\ref{synergy})) of sectoral unit productivity imposition in terms of log price reduction in the restoring CCES economy. 
Right: Synergy effect in the Leontief economy. 
}
    \label{fig_synergy}
\end{figure}

Before turning to the results, let us consider how these productivities translate into reduced equilibrium prices.
Moreover, we are concerned about wheter the sum of the resulting effects (on price reduction) of independent impositions may differ from the resulting effect of simultaneous imposition.
We say that there is synergy if simultaneous imposition is more effective than the aggregative effect of independent imposition.
Accordingly, we define synergy in terms of the log price reduction as follows:
\begin{align} 
\text{Synergy} 
= - \ln \mathcal{E}\left( e^{\sum_{j=1}^J \ln \check{\bm{\tau}}_1(j)}, r_1, w_1 \right) + \sum_{j=1}^J \ln \mathcal{E}\left( \check{\bm{\tau}}_1(j), r_1, w_1 \right)
\label{synergy}
\end{align}
where $\sum_{j=1}^J\ln \check{\bm{\tau}}_1(j) = \ln \check{\bm{\tau}}_1(\text{all})$ defines simultaneous imposition.
Notably, both terms on the right-hand side reduce to $\ln \check{\bm{\tau}}_1(\text{all}) \left[  \mathbf{I} - \mathbf{A} \right]^{-1}$ in the Cobb-Douglas economy, in which case there are zero synergies.
Otherwise, synergy can be significant.
Figure \ref{fig_synergy} displays synergies in the restoring CCES and Leontief economies.
We observe positive synergy for the restoring CCES economy, whereas the synergy is relatively small and negative for the Leontief economy.
In other words, simple summation of independent estimates of sector-wise productivity changes can underestimate the economy-wide effect of simultaneous sectoral productivity changes in the restoring CCES economy, while these economy-wide effects can be overestimated in the Leontief economy.

\begin{figure}[t!]
\centering
    \includegraphics[width=0.495\textwidth]{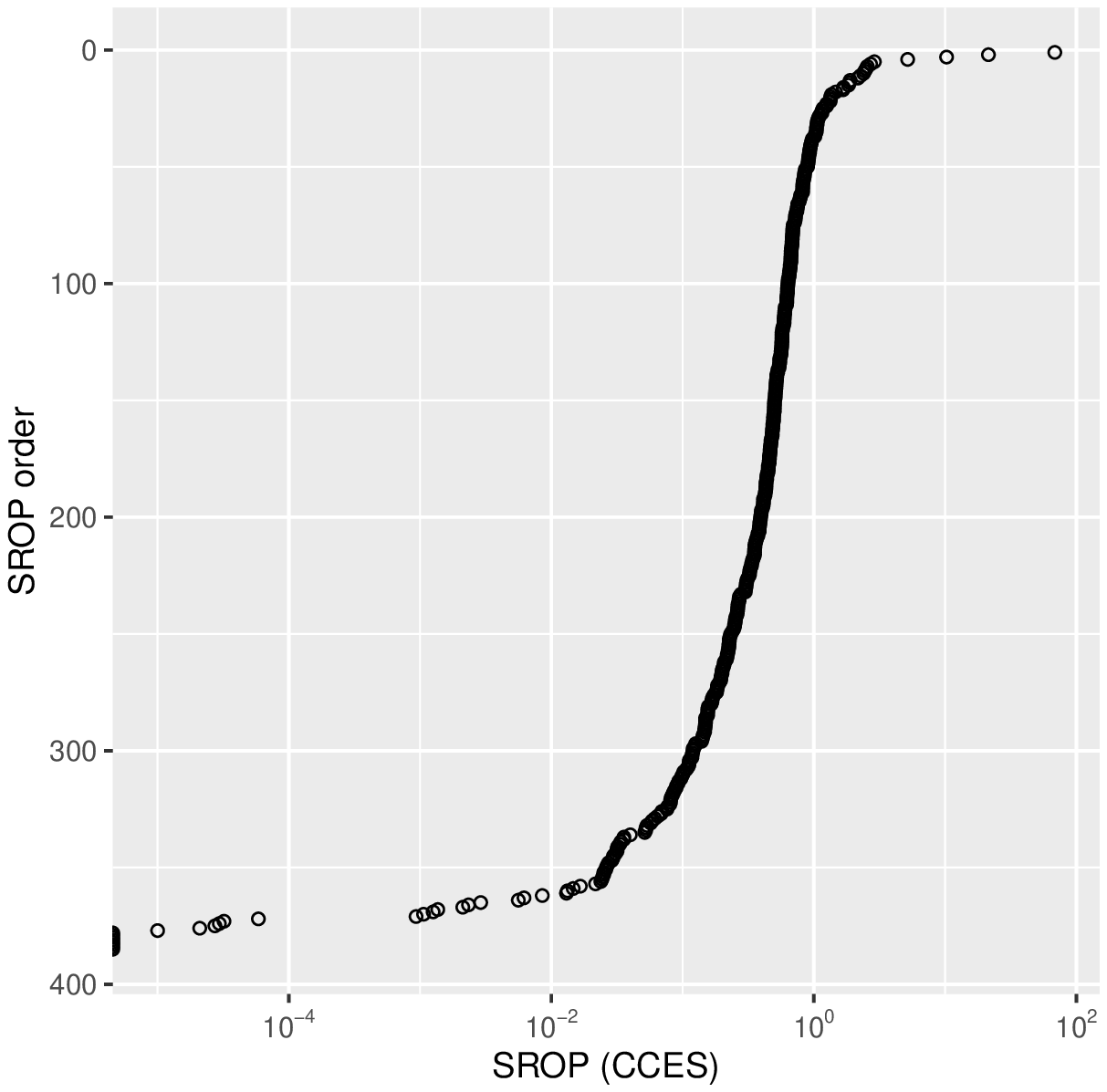}
    \includegraphics[width=0.495\textwidth]{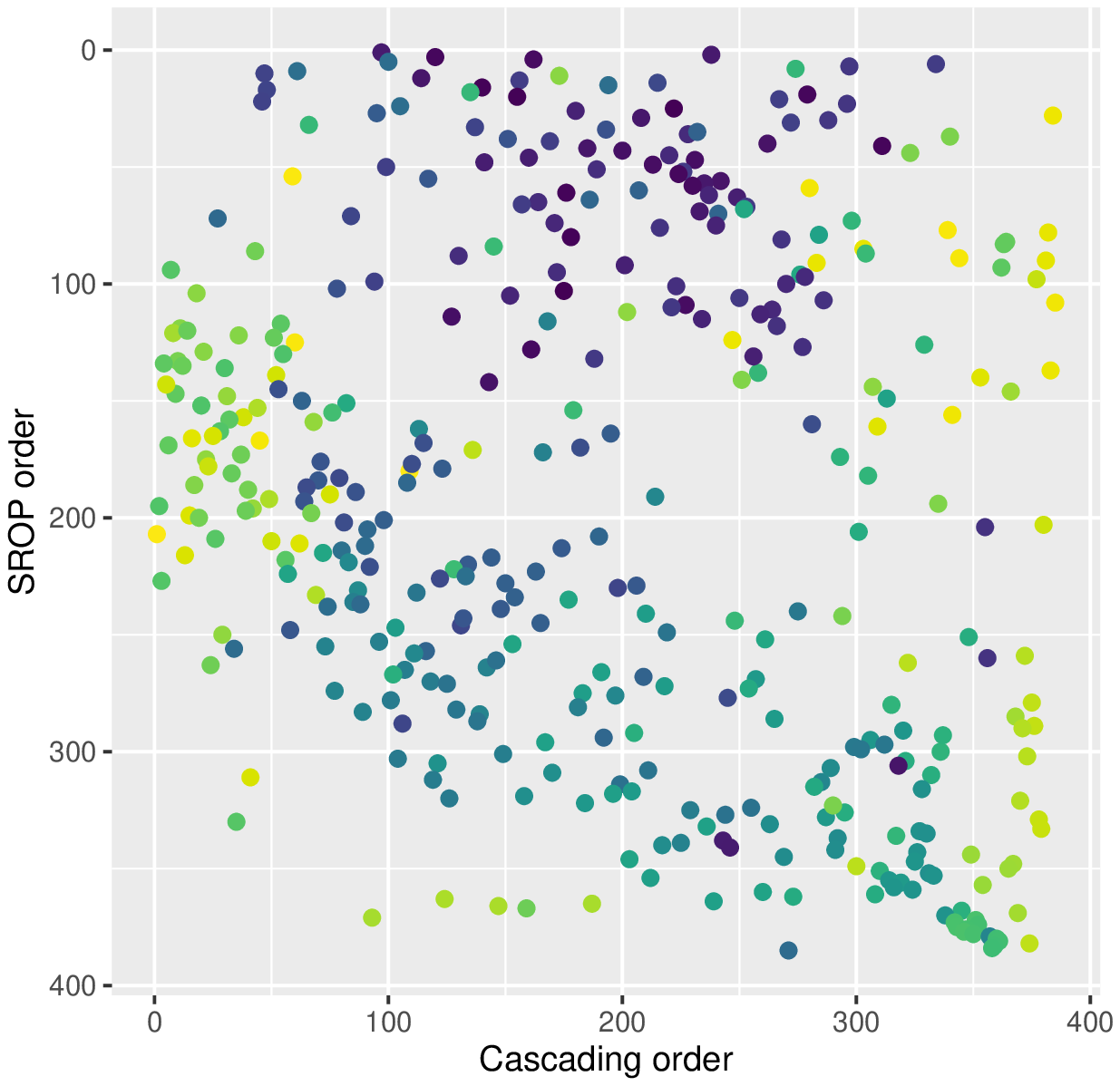}
    \caption{
Left: SROP (defined as (\ref{srop}) for all sectors.
Right: Correspondences between SROP order and cascading order.
}
    \label{fig_indep}
\end{figure}

Figure \ref{fig_indep} shows the results of independently imposing unit productivity in all sectors $j=1,\cdots,J$.
The left figure displays $\text{SROP}$ in descending order (which provides the SROP order).
The right figure displays this SROP order against the cascading order, while colors correspond to the classification order.
We observe two clusters in this figure.
Regarding the classification order, the cluster of sectors with the highest SROP (e.g., Coal mining, crude petroleum and natural gas, Metallic ores, Miscellaneous edible crops, etc.) are typically primary industries (marked with dark colors). 
Another cluster is placed on the lower left-hand side of the figure.
These are generally secondary and tertiary sectors, regarding the classification order, and regarding the cascading order, upstream (downstream) sectors have larger (smaller) SROP. 
Finally, we impose standard productivity simultaneously in all sectors, where we estimate $\text{SROP}\left(\check{\bm{\tau}}_1(\text{all})\right) = 0.727$, whereas a simple summation of the independent effects amounts to $\sum_{j=1}^I \text{SROP}\left(\check{\bm{\tau}}_1(j)\right) = 0.694$.
Hence, concerning the potential positive synergy observable in the restoring CCES economy, the underestimation can amount to ${0.727}/{0.694}-1 = 4.9\%$.

\section{Concluding Remarks}
In this study, we essentially model the \textit{metastructure} of economy-wide production that summarizes network transformation as observed in a set of linked input-output tables. 
This model allows us to study shorter scaled but detailed transformations of production networks.
As the model encompasses potential alternative technologies in each sector, it comprises a set of sectoral production functions spanning many substitutable factor inputs. 
Each sectoral production process is modeled by binary compounding processes ultimately cascaded in a universal sequence.
As we discover a self-similar hierarchical structure stylized in the empirical input-output transactions, we utilize a corresponding sequence as the fundamental and persistent structure underlying networks transformation.
For all sectoral production function, by assuming constant returns to scale, we find that the CES elasticity and share parameters for all binary compounding processes can be estimated by means of dynamic optimization.

We estimated all parameters by two-point regression such that the empirical general equilibrium model restores the production networks of the two periods.
Moreover, we measure sectoral Hicks-neutral productivity growth by the gap between the predicted unit cost and the observed output price.
Furthermore, we model the utility of a representative household by a multifactor CES aggregator function with a single substitution elasticity.
Our approach to households' expenditure shares enables the estimation of the substitution elasticity using fixed-effects regression that exploits the variety of the commodities consumed.
The shape of the representative utility was found to be essentially Leontief.
We then integrate the indirect utility function with a system of restoring CCES unit cost functions, creating a dynamic general equilibrium model that evaluates the net social benefit of a given productivity change, in light of its potential propagative effect in transforming the production networks.

Our approach provides the basis for evaluating the economy-wide propagation of productivity in terms of network transformation that is not possible in the Cobb-Douglas economy where production networks endure in the presence of alternative technologies embodied within the unit elasticity production possibility frontier.
Conversely, a non-Cobb-Douglas economy (e.g., Leontief and CCES) is nonlinear in the sense that the networks do not persist unchanged following productivity changes, so that a Leontief inverse can no longer represent production networks for evaluating general equilibrium repercussions.
We study this nonlinearity of the non-Cobb-Douglas economy regarding microeconomic productivity shocks causing macroeconomic fluctuations and potential synergies in productivity and evaluating their social benefits.
CCES, as it stands, may offer considerable scope for modification, although its versatility should lead to a variety of applications in modeling the economy's metastructure.

\clearpage
\section*{\hypertarget{app1}{Appendix 1}: Cascading Order}
\begin{figure}[h!]
\centering
\includegraphics[scale=1]{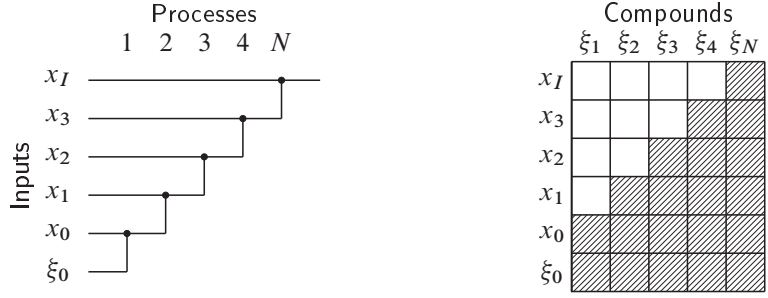}
 \caption{Cascading configuration of a production (left) and the corresponding incidence matrix (right) spanning direct and indirect inputs and intermediate outputs.\label{snc}}
 \end{figure}
Consider a cascaded (serially nested) production system comprising ${i}$ binary processes compounding ${I}+2$ inputs ($I$ intermediate and 2 primary) in ascending order indexed by $n=i+1$ \citep[see, ][for more details]{nn}.
Define incidence such that $\phi_{in} =1$ if compound input $i$ enters process $n$ directly or indirectly and $\phi_{in}=0$ if $i$ never enters process $n$ even indirectly.
For the case of cascaded production, the incidence matrix ${\Phi} =\left( \phi_{in} \right)$ becomes triangular, i.e., $\phi_{in}=1$ iff $i < n$ and $\phi_{in}=0$ iff $i \geq n$.
Furthermore, every process $n=1,\cdots,{i}$ constitutes part of an overall sequence of the compounding processes. 
That is, compound product $n$ is produced by the two primary and compound inputs $k=1,\cdots,i=n-1$, in this order.
Given that the underlying production is binary compounding, the processing sequence unravels if the ordering of inputs (or binary processes) makes the incidence matrix triangular.\footnote{To this end, however, any circular flow must be ruled out.}

Let us now focus on the $k$th process of $N$ cascading production processes.
Define $\sum_{i=0}^{I} \phi_{ik}$ and $\sum_{n=1}^{N} \phi_{kn}$ as the \textit{indegree} and \textit{outdegree} of the $k$th process, respectively.
For a perfectly triangular incidence matrix $\Phi$, the indegree-outdegree ratio of the $n$th process will be evaluated as follows:\footnote{\citet{cw} used the same criteria (ratios between indegree and outdegree) for categorizing industrial sectors, except that they used input coefficients $a_{ij}$ instead of incidents $\phi_{ij}$. 
For similar purposes, \citet{antrasGVC} applied the concept of average propagation length.
}
\begin{align}
\text{Indegree/outdegree of $k$}
= \frac{\sum_{i=0}^{I} \phi_{ik}}{\sum_{n=1}^{N} \phi_{kn}}
= \frac{k}{N-k+1} ~\text{ (for a perfectly triangular $\Phi$)}
\end{align}
In addition, it is convenient to use the following \textit{ranking index} to indicate the $n$th rank of $N$ alternatives:
\begin{align}
\text{Ranking index of $k$}
= \frac{N-k+1}{N}
\end{align}
Sorting ${i}$ observed values in ascending order and plotting against the ranking index gives the complementary cumulative density function (CCDF) of the observed values (with equal probability).

\begin{figure}[t!]
\centering
    \includegraphics[width=0.495\textwidth]{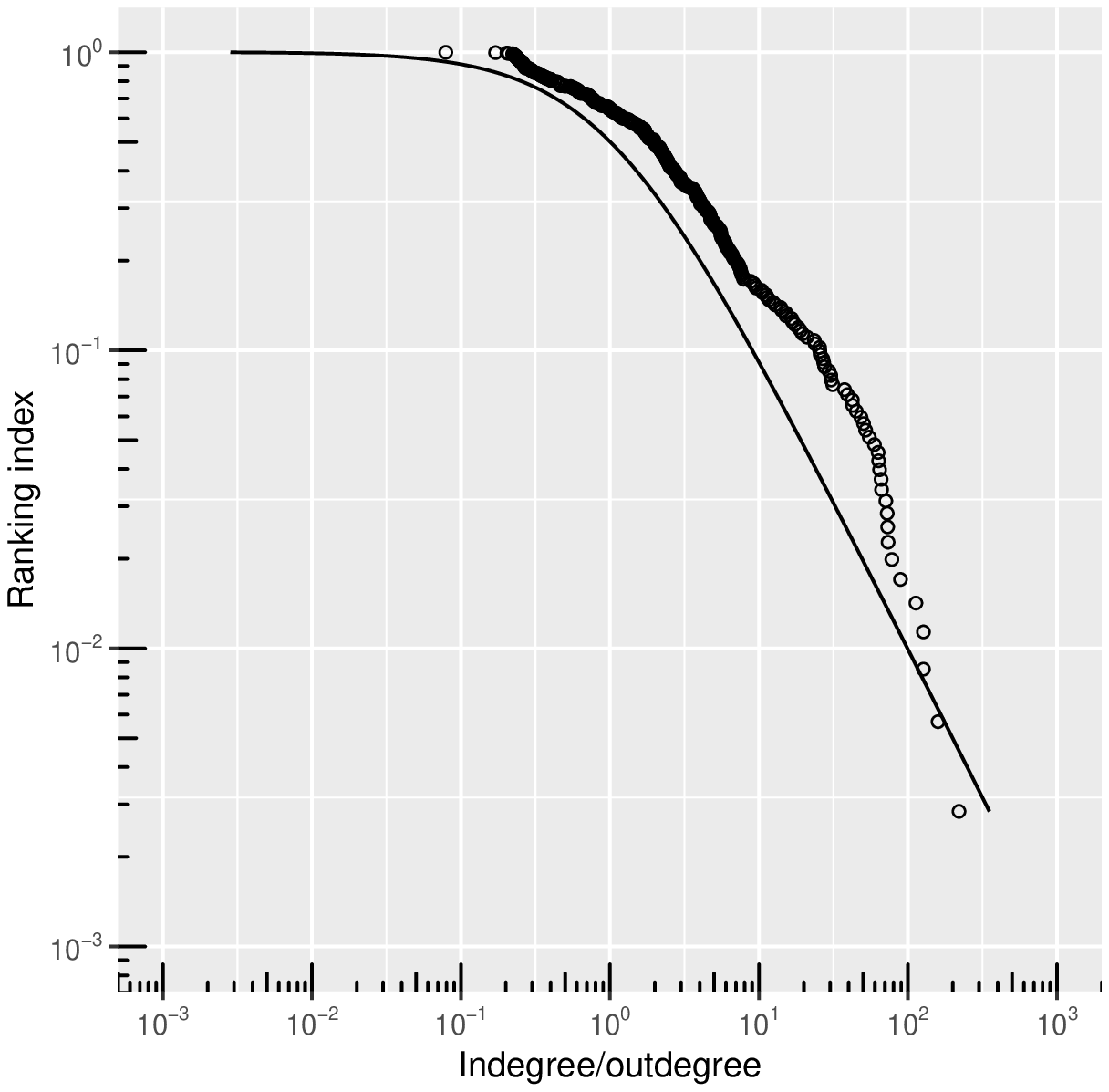}
    \includegraphics[width=0.495\textwidth]{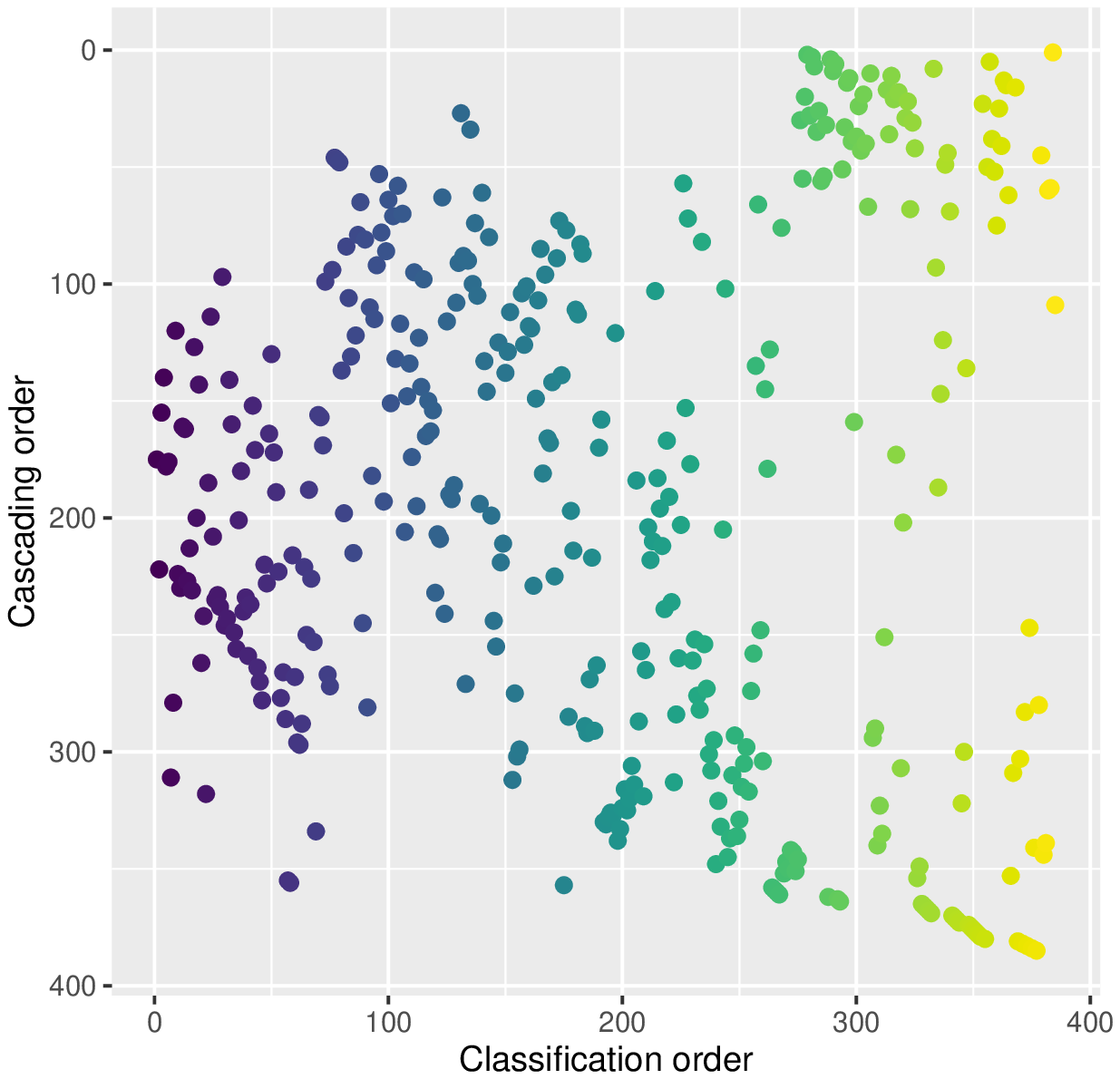}
    \caption{
Left: Open dots correspond to the CCDF of indegree/outdegree values of the 2005 input-output incidence matrix of Japan. The solid line is the CCDF of indegree/outdegree values of a perfectly triangular incidence matrix. 
Right: Low indegree/outdegree values correspond to upstream (at the top) of a stream order.  
The classification order is based on Colin Clark's primary (1--32), secondary (33--263), and tertiary (264--385) classifications. 
}
    \label{fig_order}
\end{figure}

In Figure \ref{fig_order} (left), we plot, in a solid line, the ranking index of indegree/outdegree values of an incidence matrix $\Phi$ representing cascading production (which should be perfectly triangular). 
In this case, the indegree/outdegree values of the $k$th process must be ranked $k$th in the ranking index.
We may observe linearity between the log of the two functions as $k$ approaches $N$, i.e., $\log \frac{{N}-k+1}{{N}} \approx - \log \frac{k}{{N}-k+1}$, indicating asymptotic power-law relationships between them.\footnote{In many cases a power-law distribution implies scale-freeness and self-similarity \citep{ecolett}.}
In the same figure we also plot, by open dots, the indegree/outdegree values of the incidence matrix created from the 2005 input-output table of Japan (with $\phi_{ij} = 1$ iff $x_{ij}>0$ and $\phi_{ij}=0$ otherwise) in an ascending order against the corresponding ranking index.
If the ${i}$ processes spanning the entire economy were aggregated into $J$ sectors without spoiling the hierarchy of processes, the input-output table of $J$ sectors would also have to be triangular, and its ranking index would represent the economy-wide sectoral processing from upstream to downstream. 
For empirical purposes, we also apply this hierarchy, which we hereafter call the \textit{cascading order}, to all sectoral production processes.
Figure \ref{fig_order} (right) shows the correspondences between the cascading order and input-output table's \textit{classification order}, which is based on Colin Clark's three-sector model.

\section*{\hypertarget{app2}{Appendix 3}: Substitution Elasticities of CCES}
We first examine the elasticities of a cascaded function between different factor inputs.
We begin by taking the partial derivative of a CCES aggregator with respect to $p_{i}$ and $p_{j}$ where we assume that ${i}>{j}$:
\begin{align}
\frac{\partial C}{\partial p_{i}}
&=\frac{\partial C}{\partial \pi_{I}} \cdots 
\frac{\partial \pi_{k+1} }{\partial \pi_{k}} 
\frac{\partial \pi_{k} }{\partial \pi_{k-1}} \cdots
\frac{\partial \pi_{i+1} }{\partial p_{i}} \\
\frac{\partial C}{\partial p_{j}}
&=\frac{\partial C}{\partial \pi_{I}} \cdots 
\frac{\partial \pi_{k+1} }{\partial \pi_{k}} 
\frac{\partial \pi_{k} }{\partial \pi_{k-1}} \cdots
\frac{\partial \pi_{i+1} }{\partial \pi_{i}}
\frac{\partial \pi_{i} }{\partial \pi_{i-1}} \cdots  
\frac{\partial \pi_{j+1} }{\partial p_{j}}
\end{align}
Further differentiating partially by $p_k$ where $k>i>j$ yields the following:
\begin{align}
\frac{\partial^2 C}{\partial p_i \partial p_k}
&=\frac{\partial}{\partial p_k}\left( \frac{\partial C}{\partial \pi_{I}} \cdots \frac{\partial \pi_{k+1} }{\partial \pi_{k}} \right)
\frac{\partial \pi_{k} }{\partial \pi_{k-1}} \cdots 
\frac{\partial \pi_{i+1} }{\partial p_{i}}\\
\frac{\partial^2 C}{\partial p_j \partial p_k}
&=\frac{\partial}{\partial p_k}\left( \frac{\partial C}{\partial \pi_{I}} \cdots \frac{\partial \pi_{k+1} }{\partial \pi_{k}} \right)
\frac{\partial \pi_{k} }{\partial \pi_{k-1}} \cdots 
\frac{\partial \pi_{i+1} }{\partial \pi_{i}}
\frac{\partial \pi_{i} }{\partial \pi_{i-1}} \cdots  
\frac{\partial \pi_{j+1} }{\partial p_{j}}
\end{align}
Then, we find that:
\begin{align}
\frac{\frac{\partial^2 C}{\partial p_i \partial p_k}}{\frac{\partial C}{\partial p_{i}}}
=
\frac{\frac{\partial^2 C}{\partial p_j \partial p_k}}{\frac{\partial C}{\partial p_{j}}}
=
\left( \frac{\partial C }{\partial \pi_{I}} \cdots \frac{\partial \pi_{k+1} }{\partial \pi_{k}} \right)^{-1}
\frac{\partial}{\partial p_k}\left( \frac{\partial C }{\partial \pi_{I}} \cdots \frac{\partial \pi_{k+1} }{\partial \pi_{k}} \right)
\end{align}
The above argument depends only on the $k$th input as long as the paired input's nest is inside (i.e., $k>i, j$).
Hence, the Allen-Uzawa elasticity of substitution (AUES, denoted by $\eta^{\text{AU}}$) between the $k$th input and any input inside of the $k$th compound, such as the $i$th and $j$th, will depend only on $k$. 
That is,
\begin{align}
\frac{C}{\frac{\partial C}{\partial p_{k}}}\frac{\frac{\partial^2 C}{\partial p_i \partial p_k}}{\frac{\partial C}{\partial p_{i}}}
=\eta_{ki}^{\text{AU}}
=
\frac{C}{\frac{\partial C}{\partial p_{k}}}\frac{\frac{\partial^2 C}{\partial p_j \partial p_k}}{\frac{\partial C}{\partial p_{j}}}
=\eta_{kj}^{\text{AU}}
=\eta_k^{\text{AU}}
&&
 k>i, j
\label{aues}
\end{align}
In other words, the AUES between an input and its inner-nest inputs are the same, while those between an input and its outer-nest inputs are not necessarily the same.
In Table \ref{tab_aues}, we highlight the same AUES with the same tone for the 4-nest 5-input case.

While AUES is a multifactor generalization of the two-factor elasticity of substitution, Morishima's elasticity of substitution (or MES, denoted by $\eta^{\text{M}}$) is a multifactor generalization of the original elasticity of substitution concept.\footnote{Characteristic relations between Allen-Uzawa and Morishima elasticities of substitution are discussed in detail in \citet{standup}. }
MES can be defined via AUES as follows:
\begin{align}
\eta_{ij}^{\text{M}}=a_j \left(\eta_{ij}^{\text{AU}}-\eta_{jj}^{\text{AU}} \right)
\end{align}
Here, $a_j$ indicates the $j$th factor's cost share.
Note that while AUES is symmetrical (i.e., $\eta_{ij}^{\text{AU}}=\eta_{ji}^{\text{AU}}$ for any $i\neq j$), MES is not necessarily so.
Hence, with regard to (\ref{aues}), AUES is the same for all inner-nest inputs relative to the reference nested input.
That is, if $i>j$, then $\eta_{ij}^{\text{AU}}=\eta_{i}^{\text{AU}}$, while if $i<j$, then $\eta_{ij}^{\text{AU}}=\eta_{j}^{\text{AU}}$.
This leads to the following exposition of MES for a cascaded function:
\begin{align}
\begin{aligned}
\eta_{ij}^{\text{M}}&=a_j \left(\eta_{ij}^{\text{AU}}-\eta_{jj}^{\text{AU}} \right)
=a_j \left( \eta_{i}^{\text{AU}}-\eta_{jj}^{\text{AU}}  \right) 
&&
i>j   
\\
\eta_{ij}^{\text{M}}&=a_j \left(\eta_{ij}^{\text{AU}}-\eta_{jj}^{\text{AU}} \right)
=a_j \left( \eta_{j}^{\text{AU}}-\eta_{jj}^{\text{AU}}  \right) 
= \eta_j^{\text{M}}
&&
j>i
\end{aligned} 
\label{moes}
\end{align}
In Table \ref{tab_mes} we highlight the same MES with the same tone for the 4-nest, 5-input case.

Below, we examine the elasticities of a CCES aggregator.
Without loss of generality, we focus on the $i+1=n$th nest and write the unit cost function as follows:
\begin{align}
\pi_N = C\left( \Omega_{n} \right) 
&&
\Omega_n =\left(\pi_{n}\right)^{\gamma_{i}} 
= \alpha_{i} \left(p_{i}\right)^{\gamma_{i}} + \left( 1 - \alpha_{i} \right)\left(\pi_{i}\right)^{\gamma_{i}} 
\label{hq}
\end{align}
We hereafter use $C^\prime = \frac{\diff C}{\diff \Omega_n}$ and $C^{\prime\prime}=\frac{\diff^2 C}{\diff (\Omega_n)^2}$.
For later convenience, we note that at the last nest, ${n} =N$, the following must be true:
\begin{align}
C\left( \Omega_N \right)=(\Omega_N)^{{1}/{\gamma_I}} 
=\pi_{N}
\label{lastn}
\end{align}
The key partial derivatives for examining the elasticities between inputs ${i}$ and ${i}-1$ 
follow below:
\begin{align}
\frac{\partial C}{\partial p_{i}}
&= 
C^\prime \alpha_{i}\gamma_{i} (p_{i})^{\gamma_{i}-1} 
\\
\frac{\partial C}{\partial p_{{i}-1}}
&= C^{\prime} \alpha_{{i}-1} \left( 1 - \alpha_{i}\right) \gamma_{i} (p_{{i}-1})^{\gamma_{{i}-1} - 1} (\pi_{i})^{\gamma_{i} - \gamma_{{i}-1}}
\\
\frac{\partial^2 C}{\partial p_{i} \partial p_{{i}-1}}
&= 
C^{\prime \prime} \alpha_{i}\alpha_{{i}-1} \left( 1 - \alpha_{i}\right) (\gamma_{i})^2 (p_{i})^{\gamma_{i} - 1}(p_{{i}-1})^{\gamma_{{i}-1} - 1} (\pi_{i})^{\gamma_{i} - \gamma_{{i}-1}}
\\
\frac{\partial^2 C}{\partial p_{i}^2}
&=\alpha_{i}\gamma_{i} (p_{i})^{\gamma_{i} - 1} \left( 
C^{\prime\prime}
\alpha_{i} \gamma_{i} (p_{i})^{\gamma_{i} -1}  + C^{\prime} \left(\gamma_{i} -1 \right) (p_{i})^{-1} \right) 
\end{align}
The AUES of ${i}-1$ with respect to ${i}$ for a cascaded CES function can thus be evaluated as follows:
\begin{align}
\eta_{{i}-1 \, {i}}^{\text{AU}} 
=
\frac{C}{\frac{\partial C}{\partial p_{i}}}\frac{\frac{\partial^2 C}{\partial p_{i} \partial p_{{i}-1}}}{\frac{\partial C}{\partial p_{{i}-1}}}
= \frac{C}{C^{\prime}}\frac{C^{\prime \prime}}{C^{\prime}}
\label{bbaues}
\end{align}
Hence, the AUES for a cascaded CES function can vary depending on the $i$th and subsequent inner-factor prices.
However, an exception is the last nest, where (\ref{lastn}) has the following exposition:
\begin{align}
\eta_{{I}-1 \, {I}}^{\text{AU}} 
= \frac{C}{C^{\prime}}\frac{C^{\prime \prime}}{C^{\prime}}
= \frac{(\Omega_{N})^{1/\gamma_{I}}}{\frac{(\Omega_{N})^{-1+1/\gamma_{I}}}{\gamma_{I}}}\frac{\frac{\left( \frac{1-\gamma_{I}}{\gamma_{I}} \right)(\Omega_{N})^{-2+1/\gamma_{I}}}{\gamma_{I}}}{\frac{(\Omega_{N})^{-1+1/\gamma_{I}}}{\gamma_{I}}}
= 1 - \gamma_{I}
\label{auesn}
\end{align}
\begin{table}[t!]
\begin{center}
\begin{tabular}{c}
\begin{minipage}{0.5\hsize}
\begin{center}
\caption{AUES of a CCES function ($N=4$).}
\label{tab_aues}
\begin{tabularx}{0.95\textwidth}{|c|CCCCC|}
\hline
& $4$ & $3$ & $2$ & $1$ & $0$  \\
\hline
$4$ & $-$ 
& \cellcolor{orange!20}{$1- \gamma_4$}
& \cellcolor{orange!20}{$1- \gamma_4$}
& \cellcolor{orange!20}{$1- \gamma_4$}
& \cellcolor{orange!20}{$1- \gamma_4$}
\\
$3$ & \cellcolor{orange!20}{$1- \gamma_4$} & $-$ 
& \cellcolor{orange!45}{$\eta_{3}$}
& \cellcolor{orange!45}{$\eta_{3} $}
& \cellcolor{orange!45}{$\eta_{3}$}
\\
$2$ & \cellcolor{orange!20}{$1- \gamma_4$} & \cellcolor{orange!45}{$\eta_{3}$} & $-$ 
& \cellcolor{orange!70}{$\eta_{2}$}
& \cellcolor{orange!70}{$\eta_{2}$}
\\
$1$ & \cellcolor{orange!20}{$1- \gamma_4$} & \cellcolor{orange!45}{$\eta_{3}$} & \cellcolor{orange!70}{$\eta_{2}$} & $-$ 
& \cellcolor{orange!95}{$\eta_{1}$}
\\
$0$ & \cellcolor{orange!20}{$1- \gamma_4$} & \cellcolor{orange!45}{$\eta_{3}$} & \cellcolor{orange!70}{$\eta_{2}$} & \cellcolor{orange!95}{$\eta_{1}$} & $-$\\\hline
\end{tabularx}
\end{center}
\end{minipage}
\begin{minipage}{0.5\hsize}
\begin{center}
\caption{MES of a CCES function ($N=4$).}
\label{tab_mes}
\begin{tabularx}{0.95\textwidth}{|c|CCCCC|}
\hline
& $4$ & 3 & 2 & 1 & 0  \\
\hline
$4$ & $-$ & {$\eta_{43} $} & {$\eta_{42}$} & {$\eta_{41}$} & {$\eta_{40}$}\\%\hline
3 & \cellcolor{olive!20}{$1-\gamma_4$} & $-$ & {$\eta_{32}$} & {$\eta_{31}$} & {$\eta_{30}$} \\%\hline
2 & \cellcolor{olive!20}{$1-\gamma_4$} & \cellcolor{olive!40}{$1-\gamma_3$} & $-$ & {$\eta_{21}$} & {$\eta_{20}$}\\
1 & \cellcolor{olive!20}{$1-\gamma_4$} & \cellcolor{olive!45}{$1-\gamma_3$} & \cellcolor{olive!60}{$1-\gamma_2$} & $-$ 
& {$1-\gamma_1$}
\\
0 & \cellcolor{olive!20}{$1-\gamma_4$} & \cellcolor{olive!40}{$1-\gamma_3$} & \cellcolor{olive!60}{$1-\gamma_2$} & \cellcolor{olive!80}{$1-\gamma_1$} & $-$\\\hline
\end{tabularx}
\end{center}
\end{minipage}
\end{tabular}
\end{center}
\end{table}

In Table \ref{tab_aues}, we summarize AUES following (\ref{aues}) and (\ref{auesn}).
The elasticities, which are symmetrical and equal among the inputs nested inside, equal the last parameter $1-\gamma_N$ when the elasticities are evaluated with respect to the last input.
The MES of ${i}-1$ with respect to ${i}$ can be evaluated in the same manner.
\begin{align}
\eta_{{i}-1 \, {i}}^{\text{M}}  &= a_{i} \left( \eta^{\text{AU}}_{{i}-1 \, {i}} - \eta^{\text{AU}}_{nn} \right)  
=\frac{\partial C}{\partial p_{i}}\frac{p_{i}}{C}
\left( 
\frac{C}{\frac{\partial C}{\partial p_{i}}}\frac{\frac{\partial^2 C}{\partial p_{i} \partial p_{{i}-1}}}{\frac{\partial C}{\partial p_{{i}-1}}} 
- \frac{C}{\frac{\partial C}{\partial p_{i}}}\frac{\frac{\partial^2 C}{\partial p_{i}^2}}{\frac{\partial C}{\partial p_{i}}}
\right)
\\
&=p_{i}\left( 
\frac{
C^{\prime\prime}
\alpha_{i} \gamma_{i} (p_{i})^{\gamma_{i} - 1}}{
C^{\prime}
}
-
\frac{
C^{\prime\prime}
\alpha_{i} \gamma_{i} (p_{i})^{\gamma_{i} - 1}  +
C^{\prime}
\left( \gamma_{i} - 1 \right) (p_{i})^{-1}  }{
C^{\prime}
}
\right)
=1- \gamma_{i}
\label{moesn}
\end{align}
Thus, the MES of a nested input with respect to a back-to-back inner-nest input is constant at the CES elasticity parameter of that nest.  
Moreover, according to (\ref{moes}), a nested input MES is the same with respect to any inner-nest input.
Hence, a nested input MES with respect to any inner-nest input is constant at the CES elasticity parameter of that nest.
In Table \ref{tab_mes} we summarize MES for a cascaded CES function.

Finally, we show that $\eta^{\text{M}}_{10}=\eta^{\text{M}}_{01}=1-\gamma_1$.
Below is a list of the partial derivatives we use to assess the MES for the nest at the core, i.e., $n=1$:
\begin{align}
\frac{\partial C}{\partial p_1}
&= C^\prime \alpha_{1}\gamma_1 (p_1)^{\gamma_{1}-1} 
\\
\frac{\partial C}{\partial p_{0}}
&= C^{\prime} \left( 1 - \alpha_{1}\right) \gamma_{1} (p_{0})^{\gamma_{1} - 1} 
\\
\frac{\partial^2 C}{\partial p_1 \partial p_{0}}
&= C^{\prime \prime} \alpha _{1} \left( 1 - \alpha_{1}\right) (\gamma_{1})^2 (p_{1})^{\gamma_1 - 1}(p_{0})^{\gamma_{1} - 1} 
\\
\frac{\partial^2 C}{\partial p_0^2}
&=\left(1 - \alpha_1\right) \gamma_1 (p_0)^{\gamma_1 - 1} 
\left( C^{\prime\prime} \left(1 - \alpha_1\right) \gamma_1 (p_0)^{\gamma_1 - 1}
+ C^{\prime} \left(\gamma_1 -1 \right) (p_0)^{-1} \right) 
\end{align}
Using the above terms, we acquire the following:
\begin{align}
\eta^{\text{M}}_{10}
=\frac{\partial C}{\partial p_0}\frac{p_0}{C}
\left( 
\frac{C}{\frac{\partial C}{\partial p_{0}}}\frac{\frac{\partial^2 C}{\partial p_{0} \partial p_{1}}}{\frac{\partial C}{\partial p_{1}}} 
- \frac{C}{\frac{\partial C}{\partial p_{0}}}\frac{\frac{\partial^2 C}{\partial p_{0}^2}}{\frac{\partial C}{\partial p_{0}}}
\right)
={p_{0}} \left(\frac{\frac{\partial^2 C}{\partial p_{0} \partial p_{1}}}{\frac{\partial C}{\partial p_{1}}} -\frac{\frac{\partial^2 C}{\partial p_{0}^2}}{\frac{\partial C}{\partial p_{0}}} \right) 
&=1-\gamma_1
\end{align}

\clearpage
\section*{Acknowledgements}
%The authors would like to thank the anonymous reviewers and the editor of the journal for helpful comments and suggestions.
This work was supported by JSPS KAKENHI Grant Number 16K00687 and 19H04380.
This work is implemented as part of ``Research on Potential Future Developments in Employment and Labor along with Technological Innovation, etc.'' conducted by the Japan Institute for Labour Policy and Training.
{
\raggedright 
\bibliography{bibNN}
}

\end{document}